\documentclass{aa}         

\usepackage{graphics}
\usepackage{times}

\begin{document}
 
\thesaurus{06(02.01.2;02.08.1;02.09.1;02.19.1; 03.13.708.16.6)}
   
% A&A Section 6: Form. struct. and evolut. of stars
% Accretion, accretion disks,
% Hydrodynamics,
% Instabilities,
% Shock waves,          
% Methods: numerical,
% pulsars: general
                                   
\title{Numerical Computation of Two Dimensional Wind Accretion of
Isothermal Gas}
\author{Eiji Shima\inst{1}
\and Takuya Matsuda\inst{2}
\and Ulrich Anzer\inst{3}
\and Gerhard B\"orner\inst{3}
\and Henri M.J. Boffin\inst{4}}

\offprints{U. Anzer}
             
\institute{Kawasaki Heavy Industries, Kakamigahara, Gifu, Japan
\and Department of Earth and Planetary Science, Kobe University, Kobe 657, Japan
\and Max-Planck Institut f\"ur Astrophysik, Postfach 1523, 85740
Garching, Germany
\and Department of Physics and Astronomy, University of Wales, Cardiff CF2
3YB, UK}

\date {Received / Accepted}

\titlerunning{2D wind accretion}
\authorrunning{E. Shima et al.}

\maketitle

\begin{abstract}
A new numerical algorithm for calculating isothermal wind accretion
flows has been developed and is applied here to the analysis of the
hydrodynamics of two-dimensional plane symmetric accretion flows in
wind-fed sources. Polar coordinates are used to ensure fine resolution
near the object. It is found that a thin accretion column is formed
which shows wave-like oscillations. Small accretion disks are formed
temporarily around the object. Mass accretion rate and angular
momentum accretion rate exhibit quasi-periodic oscillations. The
amplitudes of the oscillations depend on the size of the inner
boundary, the number of grid points and the method of calculation. For
a smal\-ler size of the inner boundary, finer grids and more accurate
numerical schemes, the amplitudes of the oscillation become larger. 

\end{abstract}

\section{Introduction}
Wind-fed accretion by a compact gravitating object is an important 
astrophysical phenomenon, which arises in many situations including massive 
X-ray binaries.

Wind accretion has been studied extensively both analytically and 
numerically. Hoyle \& Lyttleton (\cite{Hoy}) were the first to study an
axisymmetric wind accretion onto a gravitating object immersed in a uniform
flow. Assuming that the fluid particles collide at the axis and lose their
tangential momentum, they obtained a mass accretion rate of,
\begin{equation}
\dot{M} = \pi \rho _ \infty V _ \infty R_{\rm a} ^ 2 
\end{equation}
where the accretion radius, $R_{\rm a}$, is given by
\begin{equation}
R_{\rm a} = \frac{2 GM }{V ^2 _\infty},
\end{equation}
and $\rho_\infty$, $V _ \infty$, $G$, $M$ are the density, velocity 
of the gas at infinity upstream, the gravitational constant and the
mass of the object, respectively. In this estimate the pressure is
neglected which 
means that the Mach number of the gas is infinitely large.

Bondi \& Hoyle (\cite{bon44}) have worked on the model in more detail and Bondi
(\cite{bon52}) obtained a formula for the case of spherical accretion. The
latter
case means that the Mach number at infinity is zero. 
He also proposed an interpolation formula for flows with finite Mach number. 

The numerical computation of axisymmetric wind accretion was first attempted 
by Hunt (\cite{Hun71,Hun79}) and general agreement between the accretion
rates from 
analytical estimates and his numerical calculations was found. Shima et al.
(\cite{Shi}) 
and Koide et al. (\cite{Koi}) calculated the same problem using a finer mesh, a 
modern algorithm and a supercomputer, and found accretion rates about
a factor two larger than those obtained by Bondi.

If the flow at large distances is not uniform, angular momentum can be
transferred to the compact object causing spin up or spin down. Ho et
al. (\cite{ho}) considered this problem based on non-axisymmetric
simulations.  Two-dimensional (planar) numerical computations of an
inhomogeneous flow were carried out and non-steady 'flip-flop'  motions
were found 
by several authors (Matsuda et al. \cite{Mat87}, Taam \& Fryxell
\cite{Taa}, Fryxell \& Taam \cite{fry})

Furthermore it was found that non-steady motion occurs even in a homogeneous 
medium (Matsuda et al. \cite{Mat91}). Livio et al. (\cite{Liv}) suggested a
possible cause of the instability. \protect{Boffin} \& Anzer
(\cite{bof}) calculated two--dimensional wind accretion using a
smoothed particle method (SPH) and confirmed the
unsteadiness. Benensohn et al. (\cite{ben}) calculated uniform wind
accretion of an adiabatic gas with polytropic index $\gamma = \frac{4}{3}$
and got
'flip-flop' nonsteady motions.

Three--dimensional simulations of wind accretion were performed and compared 
with two-dimensional cases by Sawada et al. (\cite{Saw}) and Matsuda et al.
(\cite{Mat91, Mat92}). The three--dimen\-sional calculations performed
by Ishii et al. (\cite{Ish}), Ruffert (\cite{Rufa,Rufb,Rufc,Rufd}) and
Ruffert \& Arnett (\cite{Rufe}) have sufficient spatial resolution to
lead to non-steady oscillatory flows. Generally these oscillations
have lower amplitudes than the corresponding two dimensional flows.

Since the real size of any astrophysical object is usually many orders
of magnitude smaller than the accretion radius, it seems impossible
to calculate the flow all the way to the surface. Thus the inner
boundary of the computations is much larger than the real object. This
is an artifact of the simulations.

But one finds that most of these calculations show non-steady
behaviours of various degrees. The flows for which $\gamma$ is close
to units tend to exhibit the largest fluctuations.

In this present investigation we want to focus on the special case of
isothermal flows. Wind accretion with $\gamma$ close to unity has
been studied by Matsuda et al (\cite{Mat91}), Boffin \& Anzer
(\cite{bof}), and Ruffert (1996). Since these flows were more violent than
those of higher $\gamma$ they seem of particular
interest. Physically they can be thought of as approximations to
configuration of low optical depth and efficient radiative losses
resulting in a temperature distribution which is almost uniform. 

In principle it would be desirable to include an energy equation
which describes all gain and loss mechanisms of the flow. But this
approach would complicate our calculations enormously. Therefore we
shall use the approximation of isothermality.

The objective of this study is to calculate the wind accretion flows
of an isothermal gas at various Mach numbers, $\cal{M}$, and to study the
unsteadiness of the accretion column and the appearance of accretion
disks around compact objects using a fine mesh and high resolution scheme. 

Here we present the results for two dimensional wind accretion of an
isothermal gas by a new high
resolution numerical scheme with fine numerical meshes.

\section{Governing Equations and Numerical Procedure }
\subsection{Governing Equations}
The governing equations are the hydrodynamic equations of an
isothermal gas with gravity. The selfgravity of the gas is neglected.
%\begin{equation}
%{\bf Q}_t +{\bf E}_x +{\bf F}_y={\bf G}
%\end{equation}
%\begin{equation}
%{\bf Q}=\pmatrix { \rho \cr \rho u \cr \rho v \cr },
%{\bf E}=\pmatrix { \rho u \cr \rho u^2 +p\cr \rho uv \cr },
%{\bf F}=\pmatrix { \rho v \cr \rho uv \cr \rho v^2+p \cr },
%{\bf G}=\pmatrix { 0 \cr -\frac{\rho GMx}{r ^ 3} \cr -\frac{\rho GMy}{r ^
%3} \cr } , r=\sqrt{x^2 + y^2}.
%\end{equation}
These equations are written in integral form as,
\begin{equation}
\int _\Omega \frac{\partial \bf Q}{\partial t} d\tau + \oint _{\partial
\Omega}\vec{F}   ds =
\int_\Omega \vec{G} d\tau,
\end{equation}
\begin{equation}
\vec{Q}=\pmatrix { \rho \cr \rho u \cr \rho v \cr },
\end{equation}
\begin{equation}\label{eq5}
\vec{F}=m {\bf \Phi} + p {\bf N}~,~
\vec{\Phi}=\pmatrix { 1 \cr u \cr v \cr }~,~
\vec{N}=\pmatrix { 0 \cr x_n \cr  y_n \cr },
\end{equation}
\begin{equation}\label{eq6}
m =\rho {\bf \Phi}\cdot{\bf N}
\end{equation}
\begin{equation}
\vec{G}=\pmatrix { 0 \cr -\frac{\rho GMx}{r ^ 3} \cr -\frac{\rho GMy}{r ^
3} \cr } , r=\sqrt{x^2 + y^2},
\end{equation}
where  $\rho , u , v , p $ represent density, velocity in x-y direction and
pressure,$d\tau$ and $ds$ are the differential volume and surface elements
and ($x_n,y_n$) is the  outward normal of the surface element. The
pressure for an isothermal gas is defined by.
\begin{equation}\label{eq8}
p=\rho c^2,
\end{equation}
where $c$ is the constant sound speed.

The two momentum equations can be also written in terms of the angular
momentum and of the momentum in the local radial direction, leading to
%\begin{equation}
\begin{eqnarray}
\label{AMC1}
\lefteqn{\int _\Omega \frac{\partial}{\partial t} (y \rho u - x \rho v)~d\tau}
\nonumber \\
& & \mbox{} + \oint _{\partial \Omega}[ y (m u + p x_n)  - x ( m v + p y_n)
 ]~ds =0,
\end{eqnarray}
%\end{equation}

\begin{eqnarray}
%\begin{equation}
\label{AMC2}
\lefteqn{\int _\Omega \frac{\partial}{\partial t} \left[\rho (u X_r + v Y_r
) \right]~d\tau} \nonumber \\
& & \mbox{} + \oint _{\partial \Omega}\left[ X_r (m u + p x_n) + Y_r ( m v
+ p y_n) \right]~ds \nonumber \\
& & \mbox{} = \int_\Omega \left[- X_r \frac{\rho GMx}{r ^3} - Y_r
\frac{\rho GMy}{r ^3} \right]~d\tau,
\end{eqnarray}
%\end{equation}
\begin{equation}
X_r = \frac{x}{r}, Y_r = \frac{ y}{r},
\end{equation}

These momentum equations are used in the angular momentum conserving (AMC)
scheme which will be described later.

\begin{figure*}
\resizebox{0.32\hsize}{!}{\includegraphics{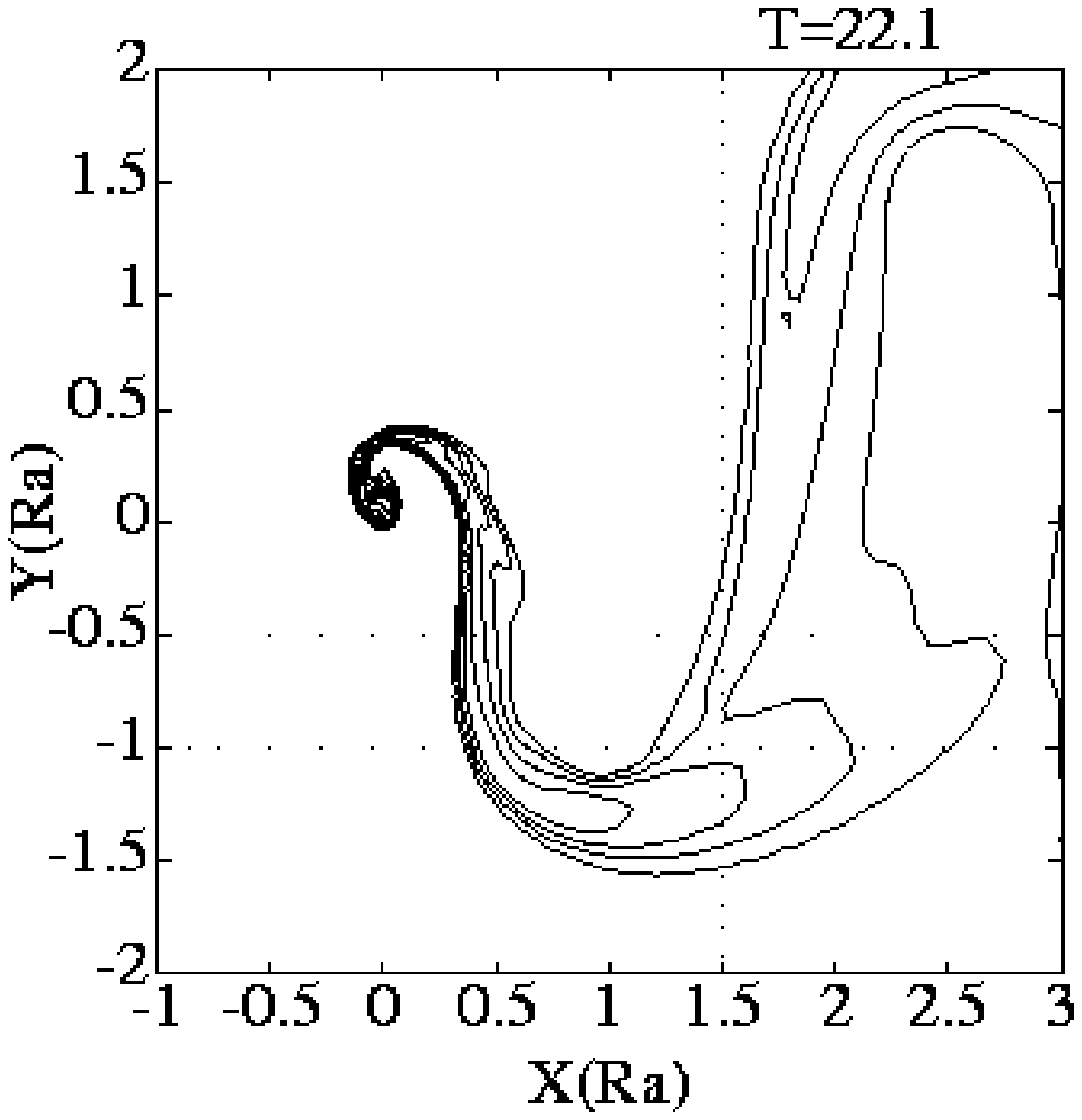}} \hfill
\resizebox{0.32\hsize}{!}{\includegraphics{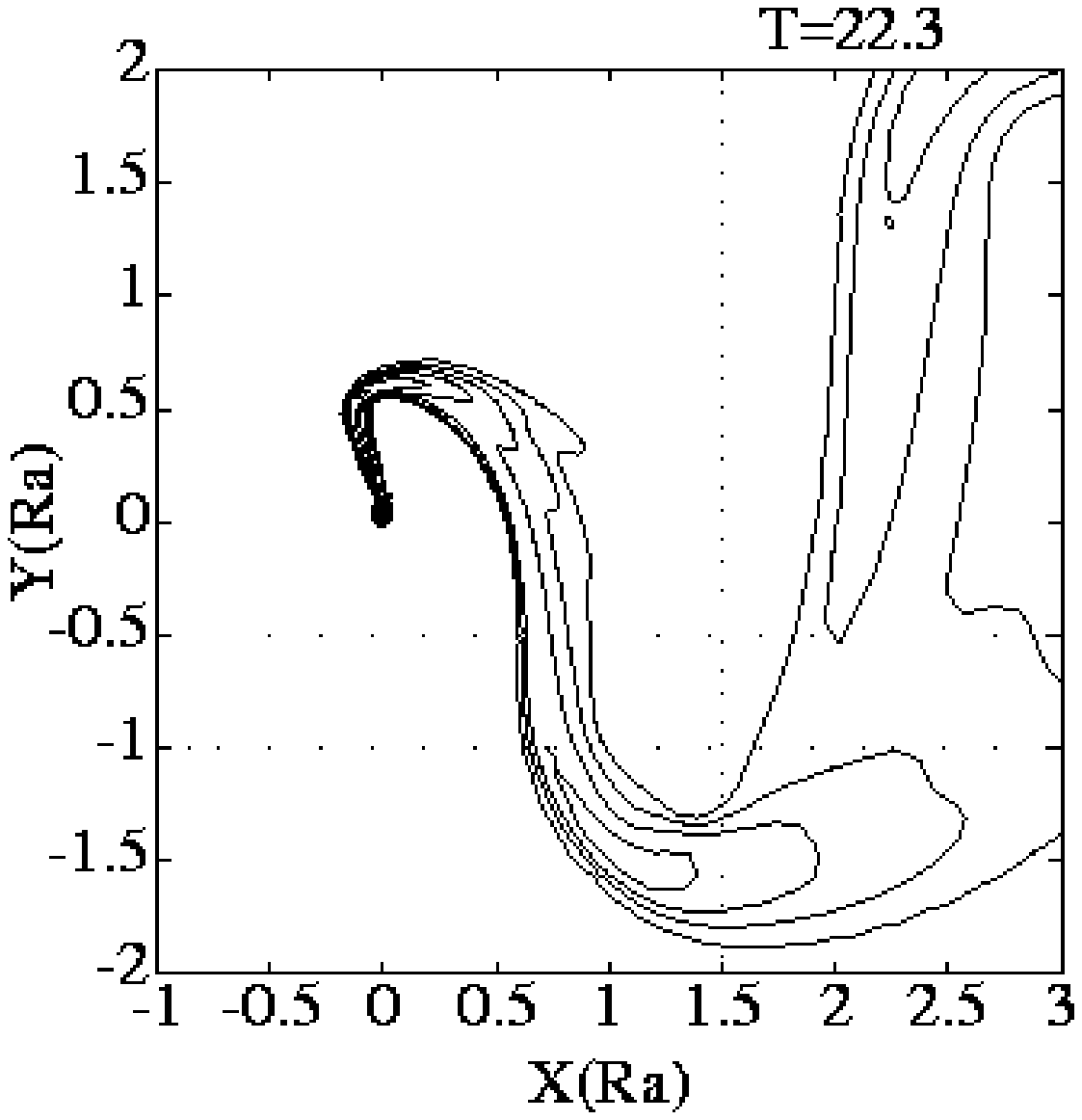}} \hfill
\resizebox{0.32\hsize}{!}{\includegraphics{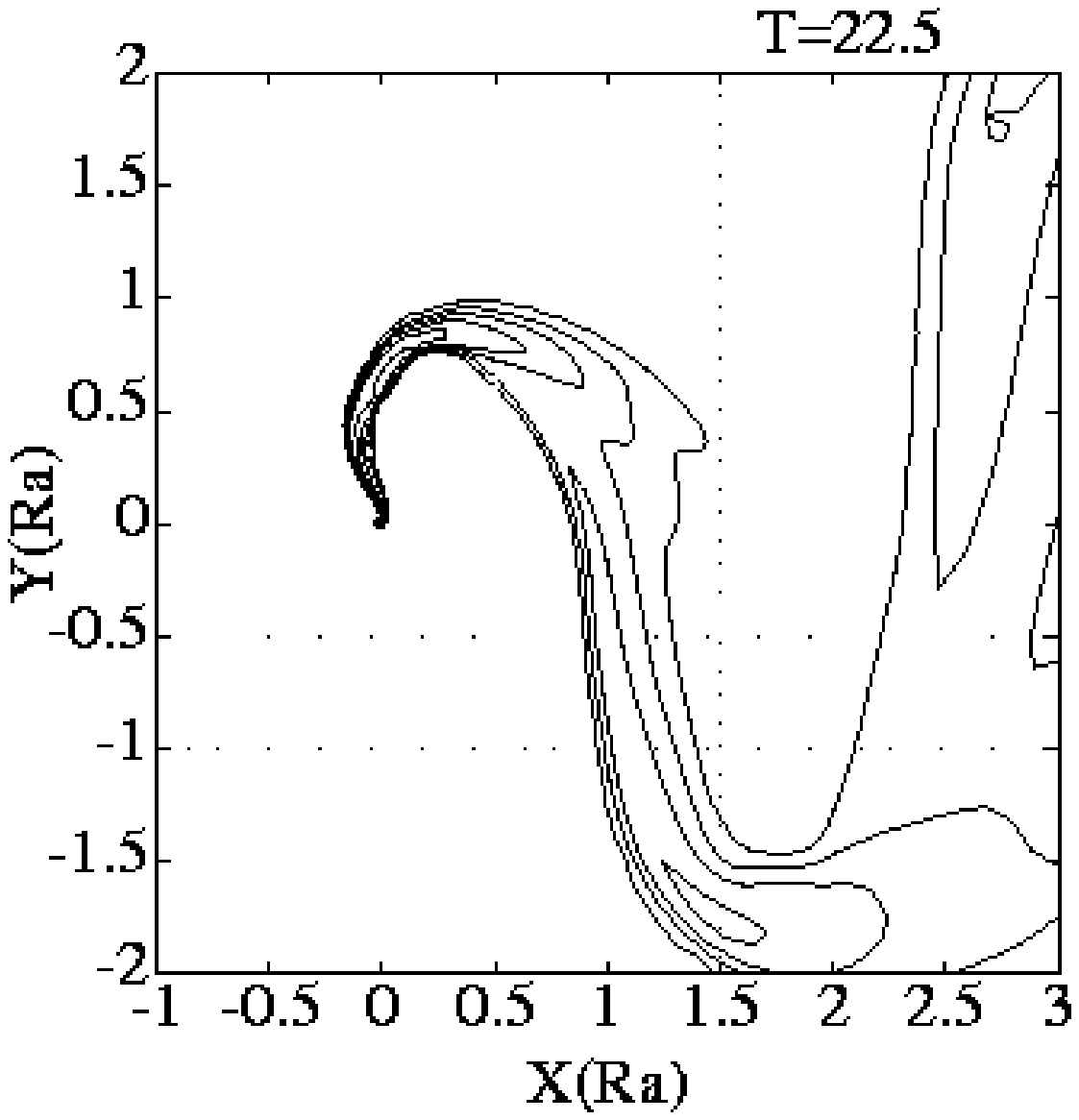}} \\
\begin{center}
\resizebox{0.32\hsize}{!}{\includegraphics{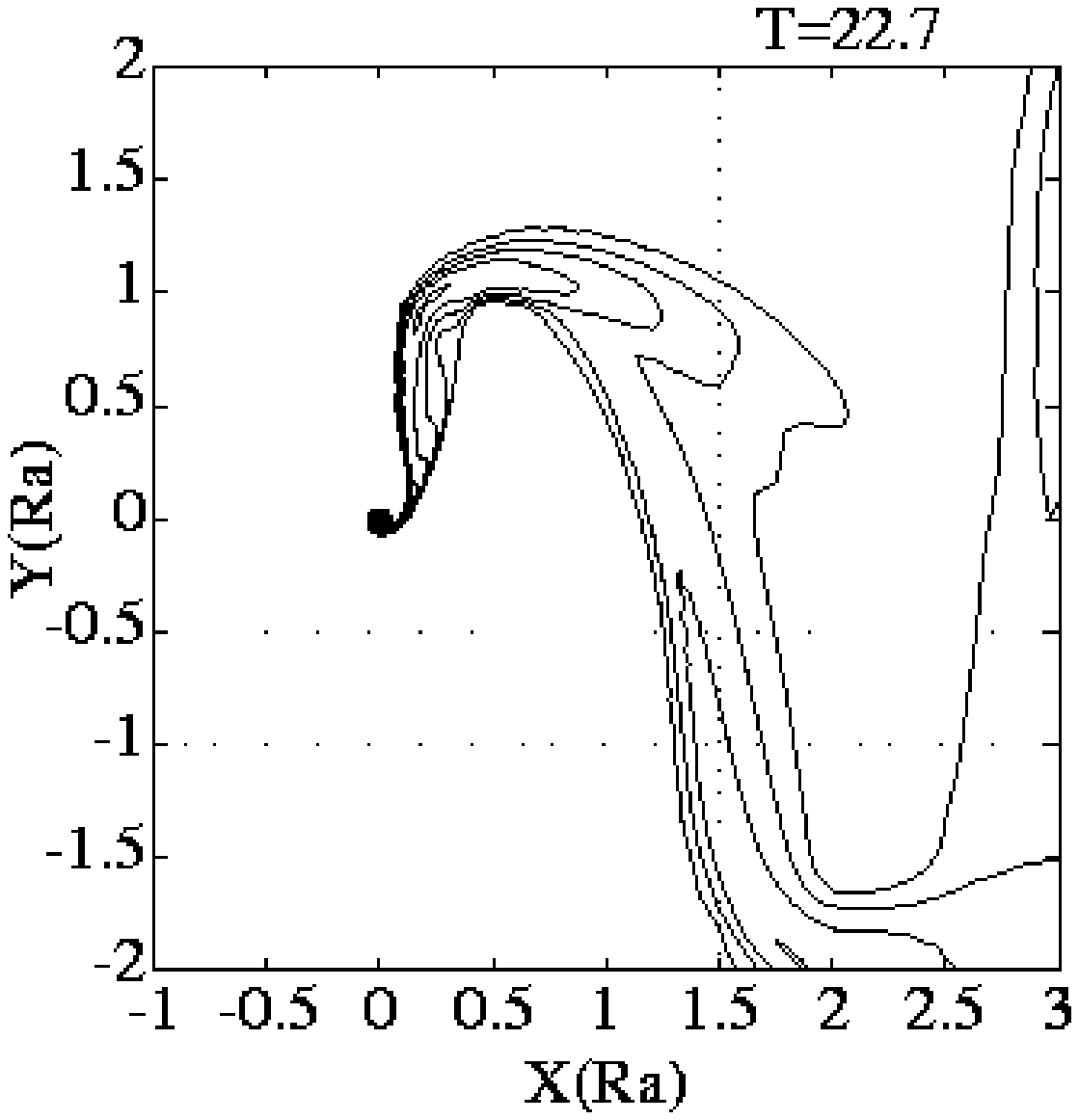}} \hspace{0.3cm}
\resizebox{0.32\hsize}{!}{\includegraphics{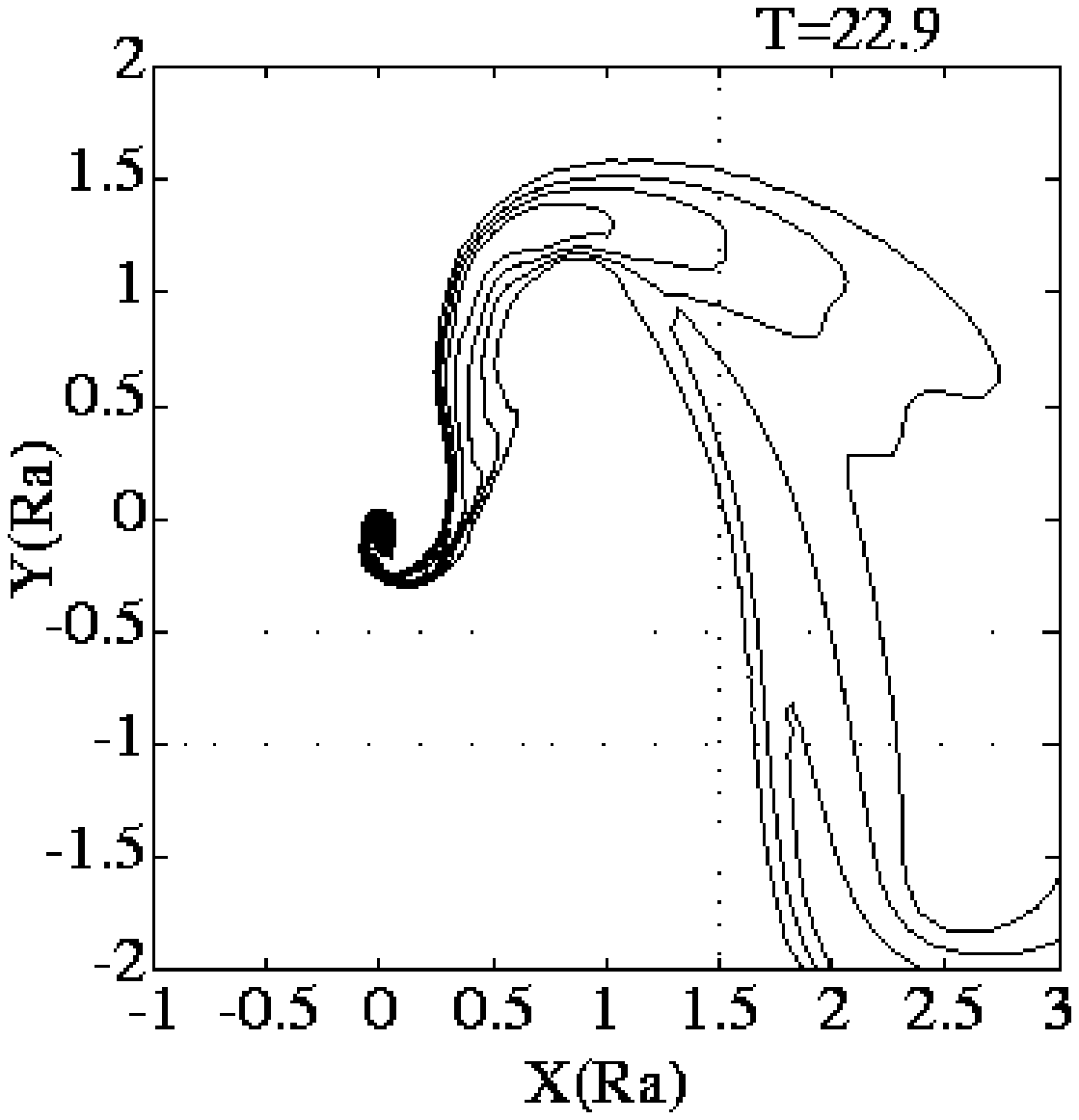}} 
\end{center}
\caption{Density contour for the AM040\_005 case, at times T=22.1, 22.3,
22.5, 22.7 and 22.9. The scale is logarithmic from 0 to 3, in steps of 0.3
.}
\label{cont222s}
\end{figure*}
 
\begin{figure*}
\resizebox{0.32\hsize}{!}{\includegraphics{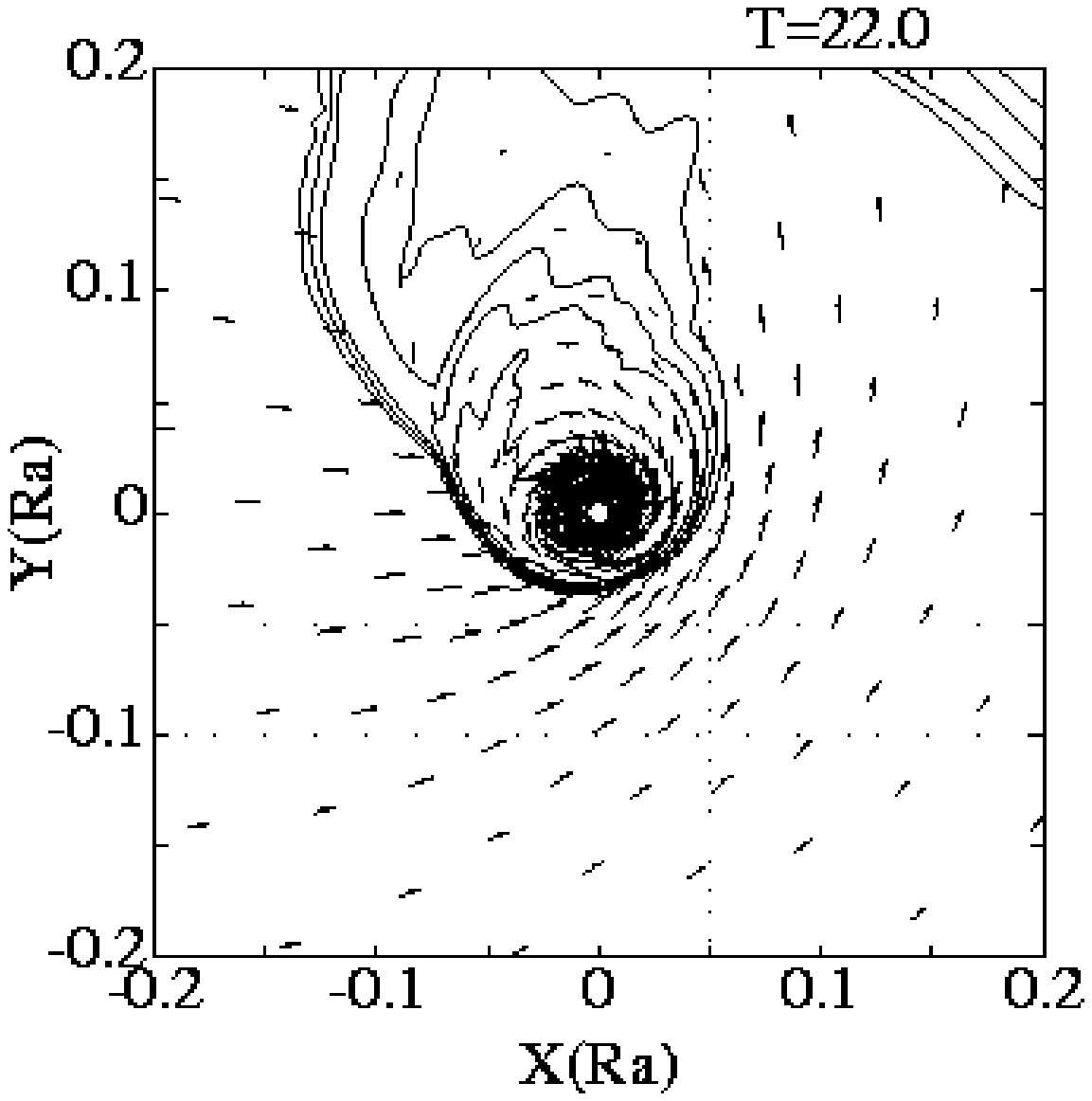}} \hfill
\resizebox{0.32\hsize}{!}{\includegraphics{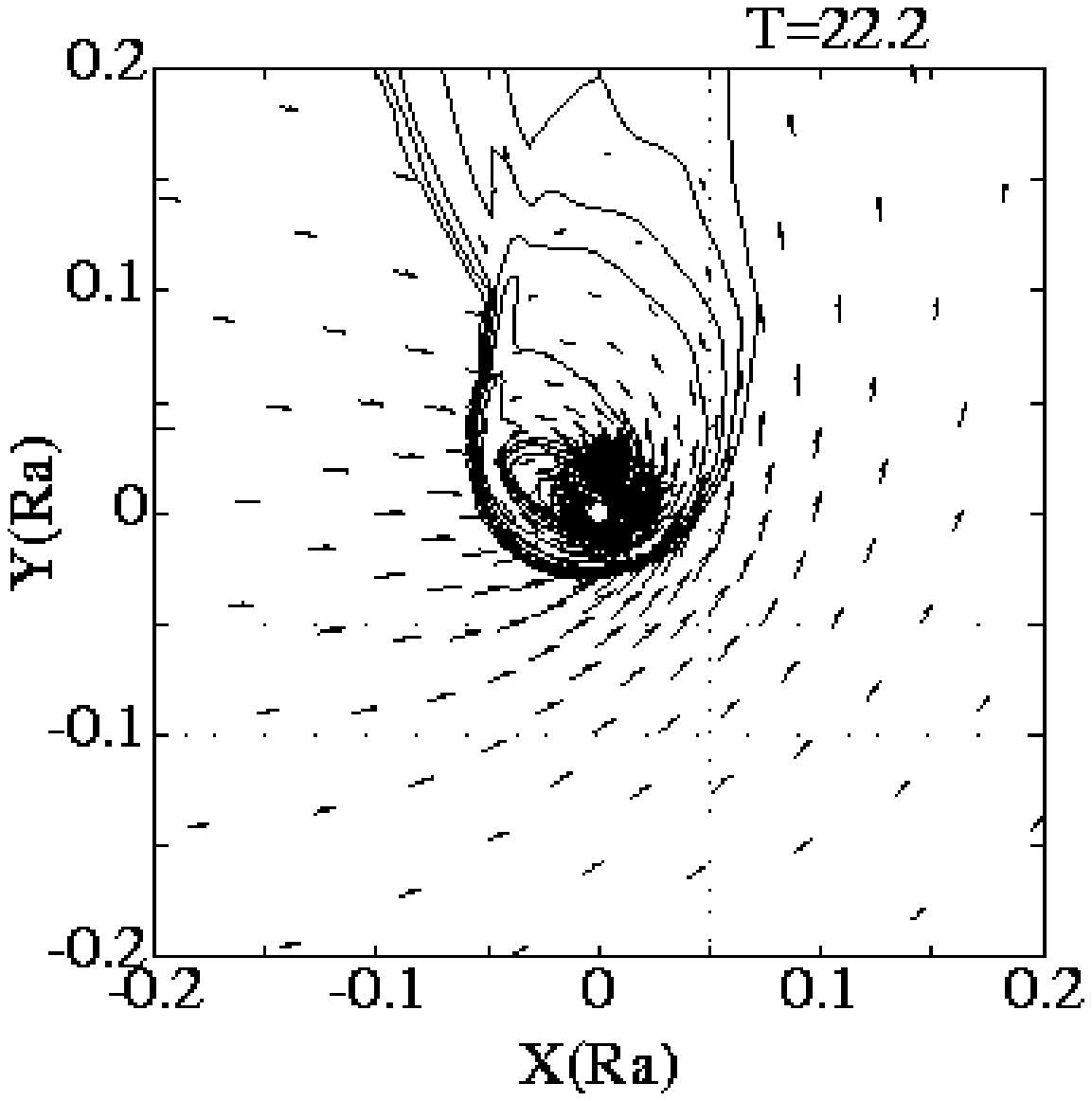}} \hfill
\resizebox{0.32\hsize}{!}{\includegraphics{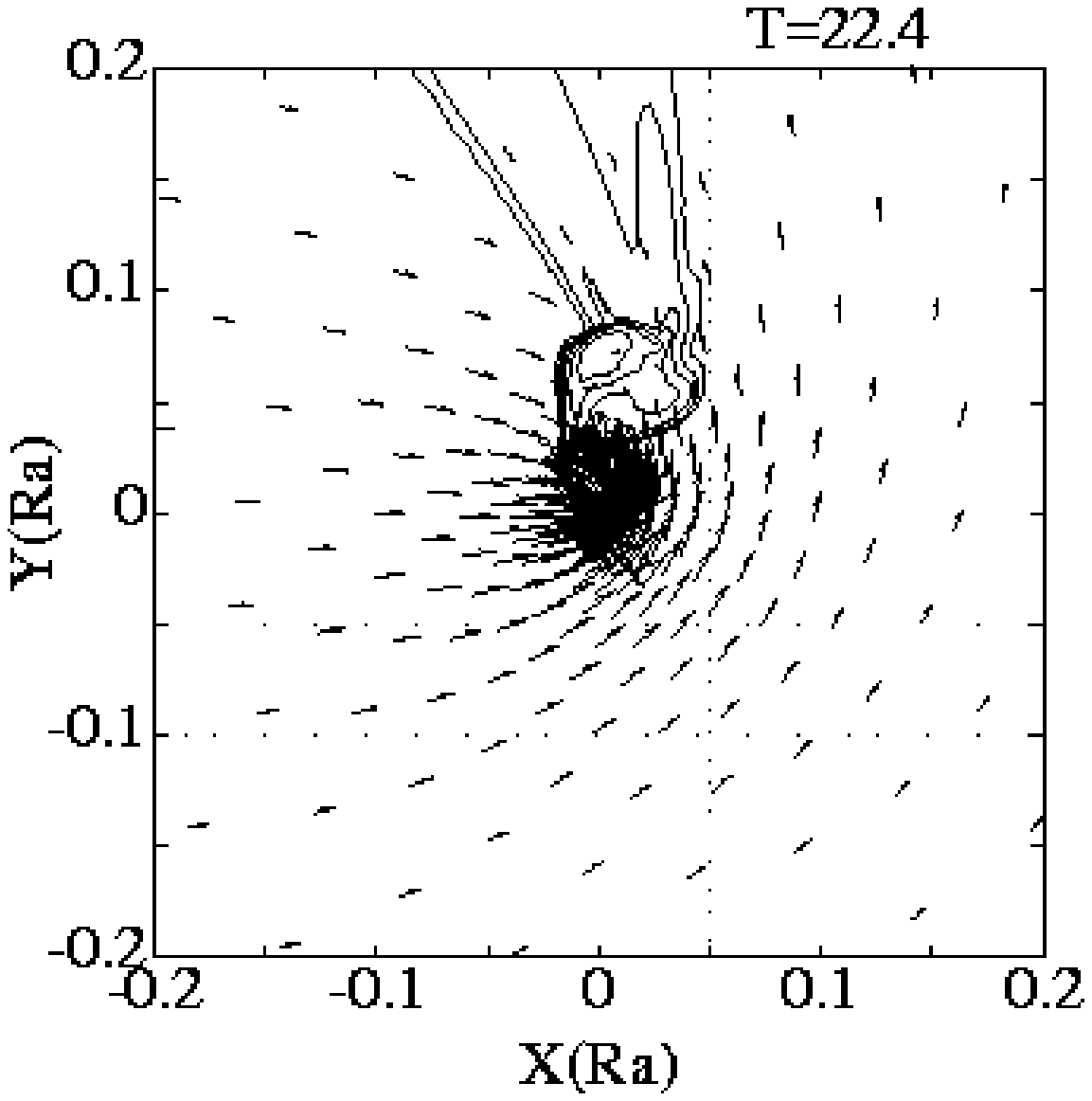}} \\
\begin{center}
\resizebox{0.32\hsize}{!}{\includegraphics{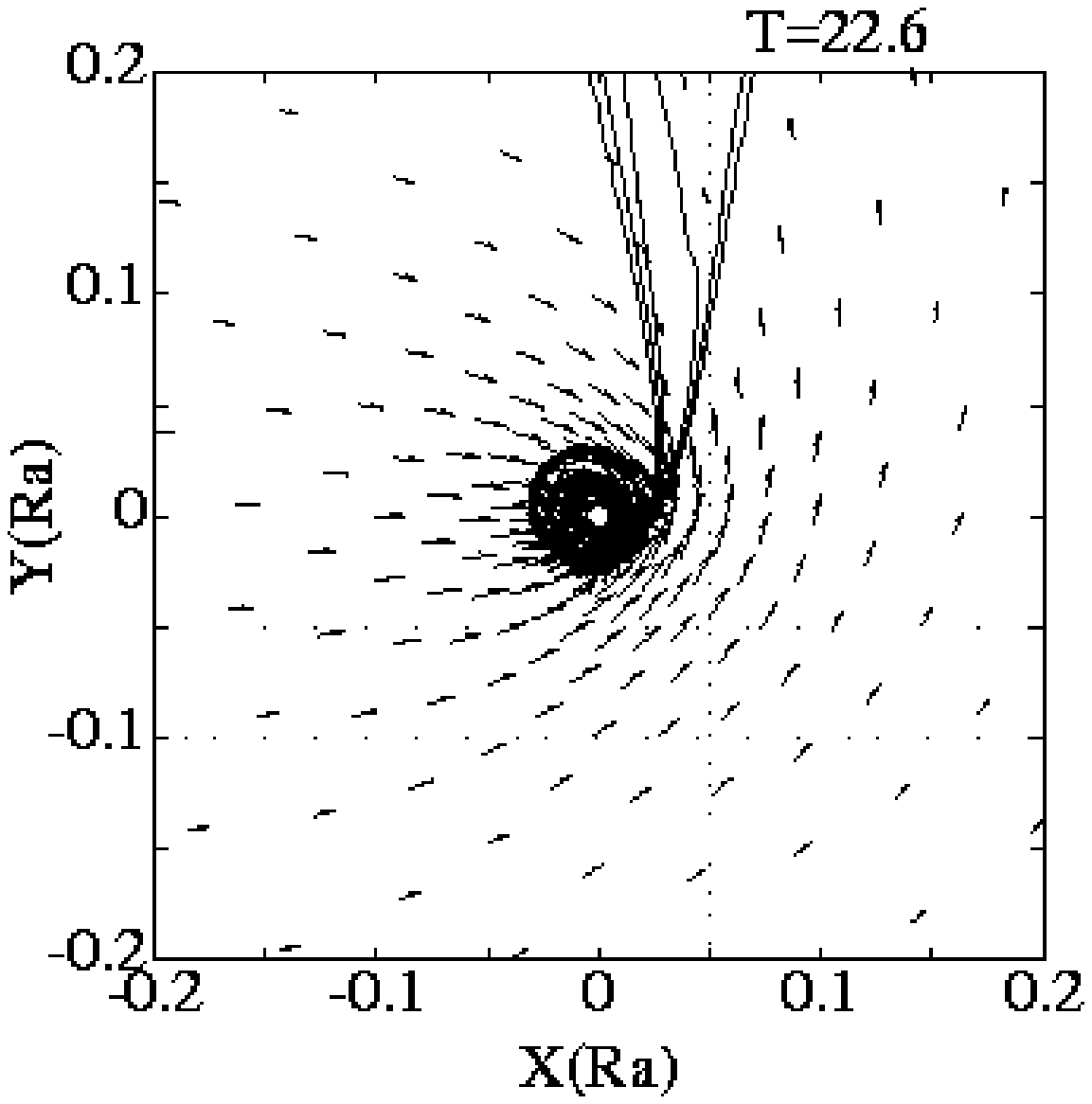}} \hspace{0.3cm}
\resizebox{0.32\hsize}{!}{\includegraphics{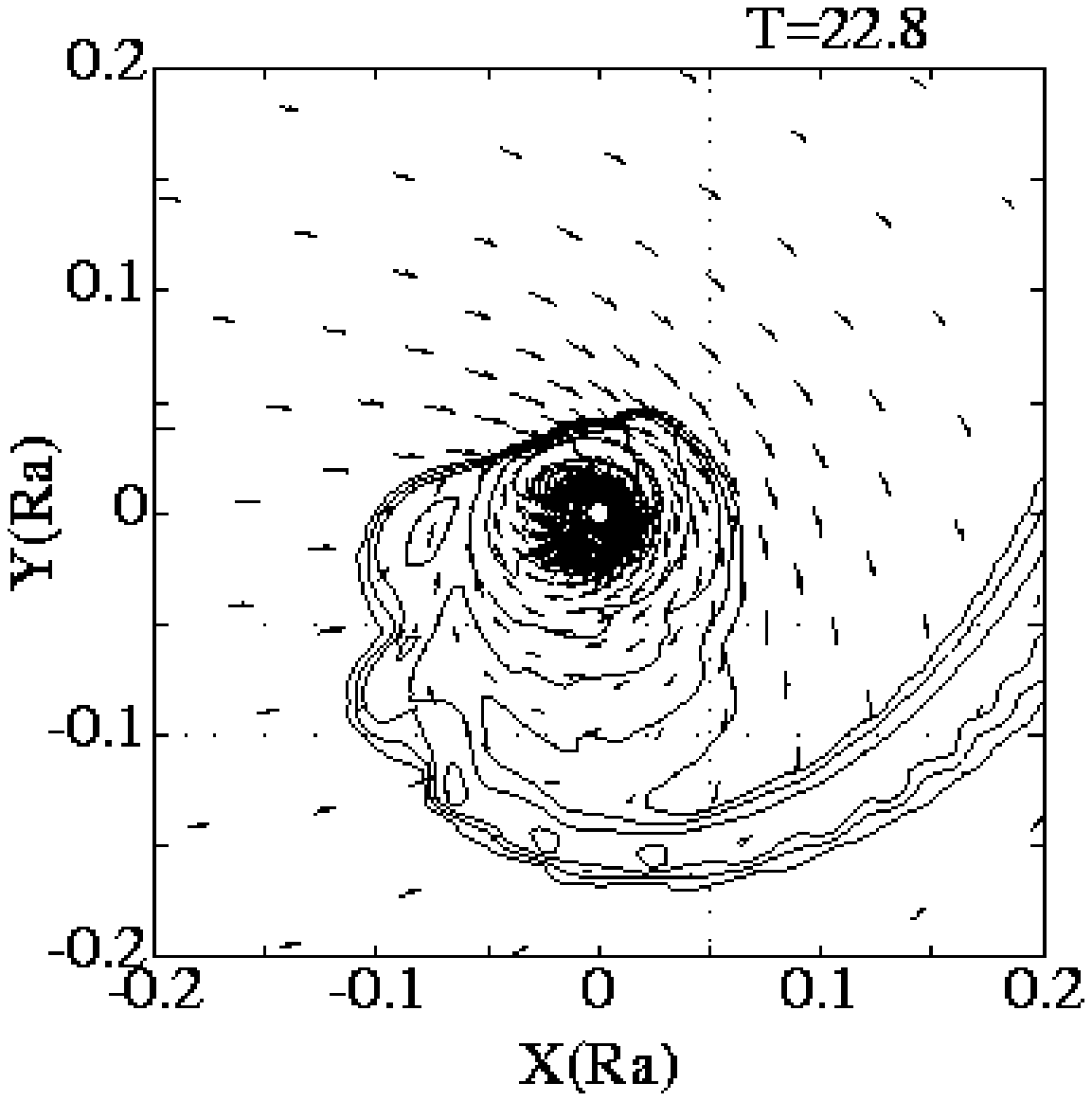}} 
\end{center}
\caption{Density contour for the AM040\_005 case, at times T=22., 22.2,
22.4, 22.6 and 22.8. The scale is logarithmic from 0 to 3, in steps of 0.6,
and we have zoomed close to the star. The arrows visualize the velocity field}
\label{cont221c}
\end{figure*}

\subsection{Initial and Boundary Conditions}
Initially the whole computational field is filled with a uniform gas. The outer
boundary condition is given by the analytic solution of Bisnovatyi-Kogan et
al. (\cite{bis}). An absorbing boundary condition, in which the density is
very low and the flow velocity is zero, is used at the inner boundary.

\subsection{Computational Mesh}
A two dimensional cylindrical grid is used for our computations.  The mesh is
divided uniformly in angular zones and is divided nonuniformly in radius.
The mesh size in the radial direction ($\Delta R$) changes exponentially 
such as,
$\Delta R_i = \Delta R_1 \alpha ^{i-1}$.

The radius of the absorbing inner boundary ($R_{\rm min}$) is 0.01 $R_{\rm a}$
and that of the outer boundary ($R_{\rm max}$) is 10 $R_{\rm a}$.  
The standard grid is made of 200 mesh points in the angular direction and
140 mesh points in the radial direction.
For the fine grid, these numbers are 400 and 250, respectively. 
The finest mesh size in the radial direction
($\Delta R_1$)  is 0.001 $R_{\rm a}$ in both the standard and the fine
grid (see Table \ref{mesh}).

Under astrophysical conditions, the inner boundary is usually located
at the magnetopause of the compact accreting object. This radius is in
general much smaller than the inner boundary of our computational
domain. But even if the two values are comparable one has to realize
that the numerical boundary conditions do not describe the physical
situation at the magnetopause which will be much more complex. The
absorbing boundary condition at the inner radius is more or less an
artificial construction. In order to study the influence of $R_{\rm
min}$, computations with different values of $R_{\rm min}$ were carried out.
\begin{table*}
\begin{tabular}{|l|c|c|c|c|c|c|}
\hline
&\ Angular Mesh & Radial Mesh & $R_{\rm max}/R_{\rm a}$ & $R_{\rm
min}/R_{\rm a}$ &$\Delta R_1/R_{\rm a} $ & $\alpha$ \\
\hline
Standard grid & 200 & 140 & 10  & 0.01 & 0.001 &1.04533\\
\hline
Fine grid & 400 & 250 & 10 & 0.01 & 0.001 &1.02200\\
\hline
\end{tabular}
\caption{Computational grids used in the simulations}
\label{mesh}
\end{table*}

\subsection{Finite Volume Method}
By applying the hydrodynamic equations in integral form to a
rectangular volume constructed by mesh lines, a finite volume
formulation is obtained. In order to obtain high spatial resolution, a
MUSCL type approach and van Albada's flux limiter are used. 
The numerical fluxes which are used by the MUSCL scheme are described
in section 2.5.

Hydrodynamic and gravitational terms are integrated simultaneously. A
two-step Runge-Kutta method is used for time integration.
This scheme is accurate to second order in both space and time. 

\subsection{Simplified Flux Splitting (SFS) for Isothermal Gas}\label{secsfs}
The finite volume method for solving the hydrodynamic equations takes
the approximation of the flux normal to the interface which are denoted by
Eqs.(~\ref{eq5}),(~\ref{eq6}) and (~\ref{eq8}).

The numerical flux in the MUSCL scheme is computed from the values defined on
both sides of the cell interface by,
\begin{equation}
\vec{F}=\vec{F}(Q _ +, Q _ -)
\end{equation}
where $\pm$ denote the values on the left($+$) and right($-$) side, and
$m$ is the mass flux. This flux is computed by solving the Riemann
initial value problem using the left and right side physical
values. It is in principle possible to solve this problem exactly, but
this would require a very large amount of computer time. Therefore approximate
but very fast algorithms have been developed.

The way in which these are calculated is vital for the MUSCL type
scheme, since it influences the computer time, robustness and
accuracy. Both the robustness for strong shock waves and the accuracy
for slip surfaces are necessary for this study, because of the high mach
number near the object and the strong shearing in the accretion column
and the disk.
 
Many algorithms for the numerical flux have been
developed. Approximate Riemann values of the fluxes were obtained by
the Roe (\cite{Roe}) and (Chakravarthy \& Osher \cite{Osher}). Although they 
are exact for shocks or rarefactions of moderate strength, they give
unphysical results for strong rarefactions or shocks
respectively. Thus they are not robust enough for hypersonic
computations like this study where very strong shocks and rarefactions
are produced by the high speed flow accelerated by gravity.
Flux vector splitting schemes (Steger \& Warming \cite{SW}, van Leer
\cite{VL}, H\"anel \& Schwane \cite{Haenel} ) are simple and more robust,
and it has been shown that they have  enough accuracy  for the shock tube
problem or flows around airfoils. However they are not accurate for contact
discontinuities such as slip surfaces, since they have excess
numerical shear stress which influences the momentum transfer especially in
accretion disks. Thus these existing schemes are not good enough for the
supersonic accretion flow. Wada \& Liou (\cite{Wada}) discussed these
problems in the context of to aerospace applications. Note that we
consider only discontinuities in one-dimension or those aligned to
grid lines in multidimensional flows. In general, discontinuities do
not align to grid lines and all schemes including exact Riemann solver
can not capture them exactly. Nevertheless it has been recognized that
those characteristics strongly affect the accuracy.

Recently Liou \& Steffen (\cite{Liou}) developed an Advection Upstream
Splitting Method (AUSM). Althogh the AUSM exhibits small overshoots at
shocks, it is as simple and robust as flux vector splitting schemes and as
accurate as the exact solver for contact discontinuities.
Inspired by  this work, a family of AUSM  type schemes have been
developed (Jounouchi et al. \cite{sfs}, Wada \& Liou \cite{Wada}, Shima \&
Jounouchi \cite{SIMA}).
It has been shown that these AUSM type schemes are sufficiently simple
robust and accurate.

Jounouchi et al. (\cite{sfs}, see also Shima \& Jounouchi \cite{SIMA}) 
developed a new AUSM type numerical flux scheme named
Simplified Flux Splitting (SFS). They reduced the
overshoot at the shock wave. Originally the SFS was developed for
adiabatic gas, but it has been modified for an isothermal gas in
this study. The SFS scheme for an isothermal gas can be written as follows.

\begin{equation}\label{flux}
\vec{F}_{SFS}= \frac{ m+|m| }{2}\vec{\Phi}_+
 + \frac{ m-|m| }{2}\vec{\Phi}_-
 + \tilde{ p }\vec{N},
\end{equation}

where $\tilde{ p }$ is an average value of the pressure defined by the
following relations:
\begin{equation}\label{pressure}
\tilde { p } = \beta _+ p_+ + \beta _- p_-,
\end{equation}

\begin{equation}
\beta _ \pm = \frac {1}{4} (2 \mp \tilde{M} _ \pm)(\tilde{M} _ \pm \pm 1)^2 , 
\end{equation}
\begin{equation}
\tilde{M}_ \pm = min(1,max(0,M_\pm)) , 
\end{equation}
and
\begin{equation}
M_\pm = \frac{V _{n \pm}}{c}.
\end{equation}

The mass flux $m$ is given by the flux vector splitting method as follows:
\begin{equation}
m = m_+ + m_- ,
\end{equation}

\begin{equation}
%\[
\left\{
\begin{array}{ll}
m_\pm = \pm 0.25 ~\rho _ \pm c  (M_\pm \pm 1)^2 & for |M_\pm|\le 1 \\
m_\pm = 0.5 ~\rho _ \pm ( V _{n \pm} \pm | V _{n \pm}| )  & for|M_\pm|> 1 .
\end{array}
\right.
%\]
\end{equation}
Note that the usage of the flux vector splitting does not degrade the accuracy
of a contact discontinuity for isothermal gas, because at the slip
surface the SFS scheme gives zero flux normal to the surface.

\subsection{Angular momentum conserving scheme}
Finite volume methods are usually based on the conservation laws of mass,
linear momentum and total energy. Conservation of energy is not used in
the present computations because the gas is isothermal. 
These conservation laws, of course, also imply the conservation of angular
momentum for the exact solutions. The conservation of linear momentum,
however, is just an approximation for the angular momentum in a
discretized description. As the accuracy is at most second order in
mesh size with any finite volume method, it depends on the numerical mesh. 

If the accreting matter in the accretion column does not collide with
itself and if it keeps some angular momentum , it will follow a Keplerian
orbit and cannot accrete. Thus the formation of an
accretion disk near the object directly reflects the angular momentum
of the accreting matter, and it therefore depends crucially on the
accuracy of the conservation of angular momentum. 

The angular momentum conserving (AMC) scheme based on Eqs.~(\ref{AMC1})
and (\ref{AMC2}) shows less mesh dependency than the linear momentum
conserving (LMC) scheme as will be shown later. The AMC scheme can 
easily be obtained from the LMC scheme. In fact, the difference between 
the AMC and LMC computer codes is only a few lines. 

\subsection{Accurate local time stepping}
The solutions of wind accretion are nonsteady in most cases. Therefore, the
numerical scheme has to be accurate in time to capture the motion
correctly. Two-step Runge-Kutta time stepping with a small Courant number (less
than 0.2) was used for this purpose. 
Near the object, the flow has very high velocity
and a fine mesh is required there for spatial
resolution, thus, the size of the time step is very restricted by the
Courant condition and very long computing times are required to reach
physically meaningful solutions, when the usual uniform step is taken. 

In order to avoid this problem, time accurate local time stepping was used in
this investigation. First a global time step is defined, then the time
step at each cell is divided into two if the Courant number is larger
than our limit. This procedure is repeated until Courant condition
is satisfied in all cells. The inner iteration is carried out in one
global iteration when the local time step is smaller than the global
time step The conservation laws are fulfilled by summing  up the
fractional fluxes correctly. A speed up by about a factor 20 was obtained
by this procedure. However, over 100 hours of computer time on a 2 GFLOPS
class supercomputer (Fujitsu VPP-300) were still needed for the fine mesh
case.

\subsection{Units}
The obvious reference length is the accretion radius $R_{\rm a}$. The fluid
velocity far upstream, $V_\infty$, is used as a reference velocity,
thus the time unit is
\begin{equation}
T_{\rm ref} = \frac{R_{\rm a}}{V_\infty}.
\end{equation}
The mass accretion rate is normalized by the  two-dimensional
Hoyle \& Lyttleton rate given by
\begin{equation}
\dot{M}_{\rm HL} = 2 \rho _ \infty V_\infty R_{\rm a}.
\end{equation}
The angular momentum accretion rate is normalized by
\begin{equation}
\dot{J}_{\rm ref} = \rho _ \infty {V_\infty}^2  R^2_{\rm a}.
\end{equation}

\begin{figure}
\resizebox{0.49\hsize}{!}{\includegraphics{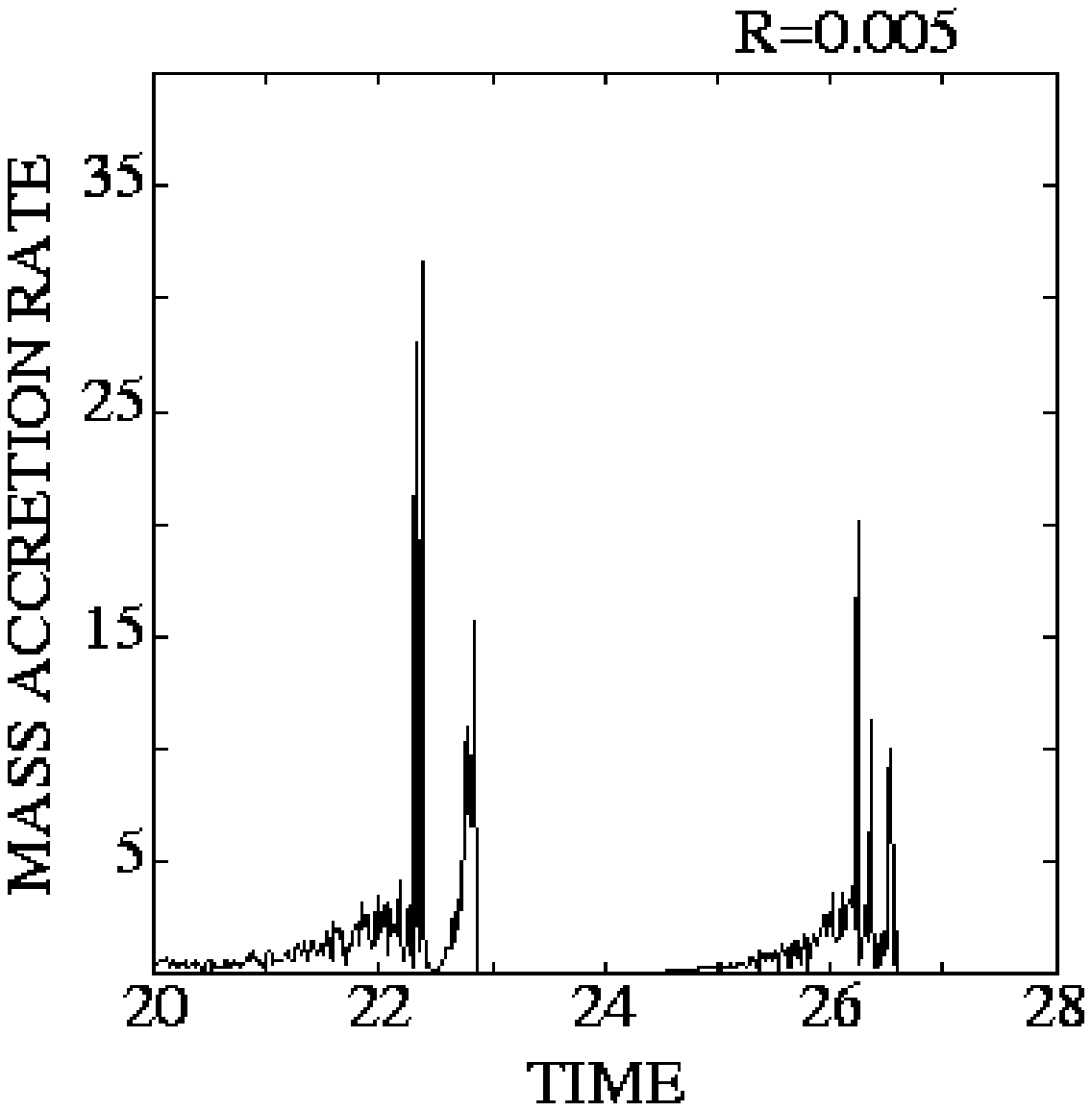}} \hfill
\resizebox{0.49\hsize}{!}{\includegraphics{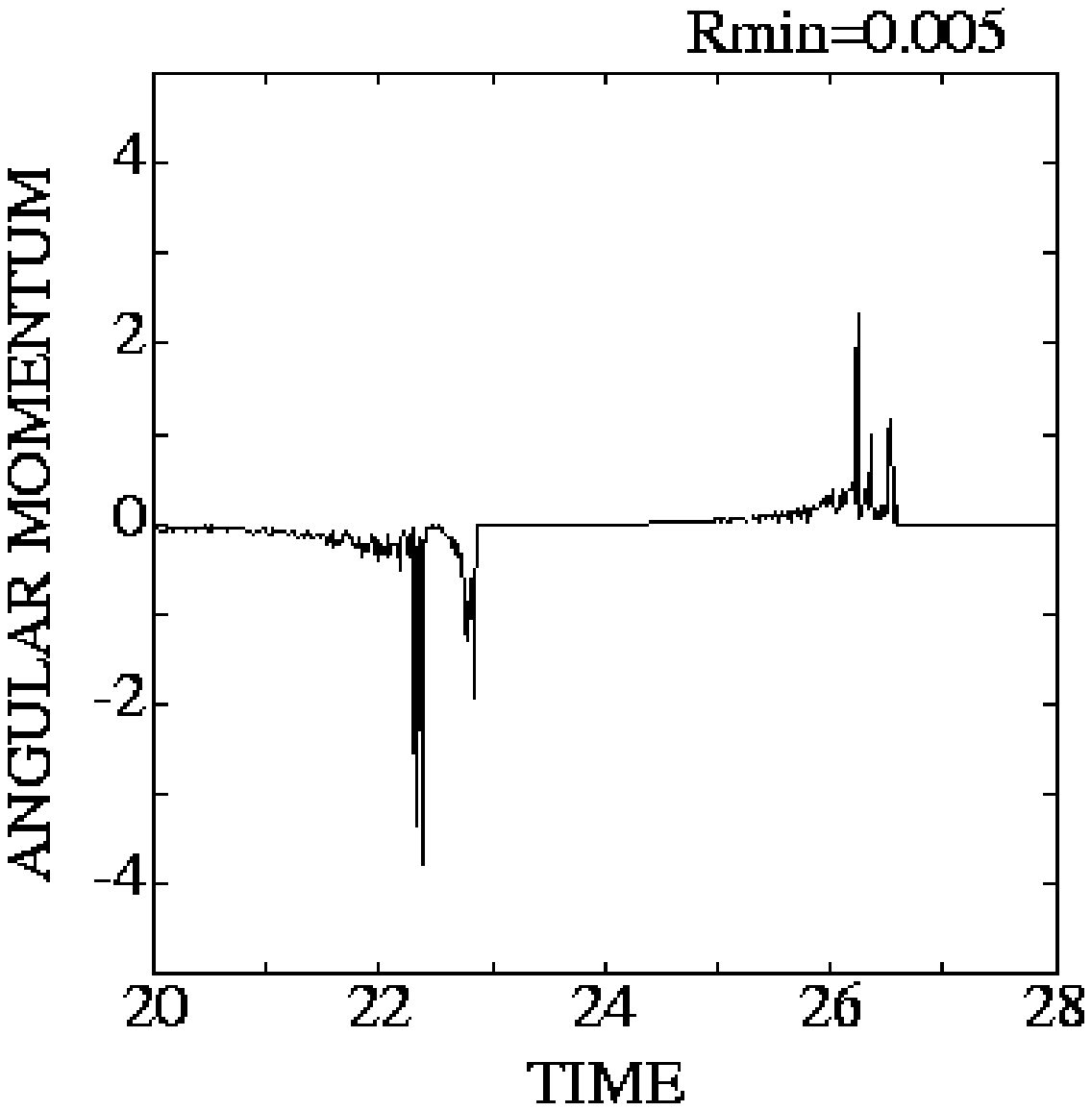}} \hfill
\caption{Time history of the mass accretion rate (left) and angular
momentum accretion rate (right) for the AM040\_005 case at times
corresponding to Fig.~\ref{cont221c}, AMC scheme}
\label{masbig}
\end{figure}

\begin{table*}
\begin{center}
\begin{tabular}{|l|c|l|r|l|r|c|l|l|c|}
\hline
& & & & & & & & & \\
Case & Scheme & ~~Grid & $\cal{M}$ & $R_{\rm min}$ & $R_{\rm max}$ &
$AVE(\dot{M})$ & ~~$\dot{M}'$ &
$AVE(\dot{J})$ & $RMS(\dot{J})$\\
\hline
\hline 
AF040\_01 & AMC & Fine & 4.0 & 0.01 & 10.0 & 0.967 & 1.331 & ~0.008 & 0.219\\
\hline
AM010\_01 & AMC & Standard & 1.0 & 0.01 & 10.0 & 1.239 & 0.092 & ~0.000 &
0.000\\
\hline
AM014\_01 & AMC & Standard & 1.4 & 0.01 & 10.0 & 0.971 & 1.623 & ~0.005 &
0.272\\
\hline
AM020\_01 & AMC & Standard & 2.0 & 0.01 & 10.0 & 0.944 & 1.961 & ~0.000 &
0.308\\
\hline
AM040\_005 & AMC & Standard & 4.0 & 0.005 & 5.0 & 0.855 & 2.031 & ~0.009 &
0.240\\
\hline
AM040\_01 & AMC & Standard & 4.0 & 0.01 & 10.0 & 0.838 & 1.176 & ~0.017 &
0.206\\
\hline
AM040\_02 & AMC & Standard & 4.0 & 0.02 & 20.0 & 0.951 & 0.786 & -0.008 &
0.201\\
\hline
AM080\_01 & AMC & Standard & 8.0 & 0.01 & 10.0 & 1.020 & 1.090 & ~0.007  &
0.160\\
\hline
AM160\_01 & AMC & Standard & 16.0 & 0.01 & 10.0 & 0.982 & 0.938 & ~0.010 &
0.125\\
\hline
\hline 
TF040\_01 & LMC & Fine & 4.0 & 0.01 & 10.0 & 1.033 & 3.040 & ~0.000 & 0.362\\
\hline
LM040\_005 & LMC & Standard & 4.0 & 0.005 & 5.0 & 0.420& 0.395 & ~0.0377 &
0.058\\
\hline
LM040\_01 & LMC & Standard & 4.0 & 0.01 & 10.0 & 0.695 & 0.643 & ~0.016 &
0.128\\
\hline
LM040\_02 & LMC & Standard & 4.0 & 0.02 & 20.0 & 0.817 & 0.683 & ~0.011 &
0.183\\
\hline
LM040\_05 & LMC & Standard & 4.0 & 0.05 & 50.0 & 0.979 & 0.3355 & ~0.0177 &
0.197\\
\hline
\end{tabular}
\caption{Parameters and nondimensional values of each case. $\dot{M}'$
is the root mean square of fluctuations from the averaged value of
$\dot{M}$. $ AVE(\dot{M})$ and $AVE(\dot{J})$ are the average of mass and
momentum accretion rate respectively, and $RMS(\dot{J})$ is the root mean
square of $ \dot{J}$.}\label{case}
\end{center}
\end{table*}

\section{Numerical Results}
\subsection{Calculated Cases}
The cases which we have calculated are summarized in table \ref{case}. 
Our main interest is in the thin accretion column of supersonic flows, thus
Mach number $\cal{M}$=4 was chosen as the standard case. The results obtained
with the LMC scheme exhibited more mesh dependency than those of the
AMC scheme, therefore we show mainly the AMC results

\subsection{Mesh Dependency of the LMC and AMC schemes}
Table \ref{case} summarizes how for both the LMC and AMC scheme the mass
accretion rate, $\dot{M}$, and the RMS value of the angular momentum
accretion rate,  $RMS(\dot{J})$ depend on the details of the
models. The mass accretion rate changes from 0.420 to 0.817 when the
size of the central hole increases from 0.005 $R_{\rm a}$ to 0.02
$R_{\rm a}$ in the LMC calculations (case LM040\_005, LM040\_01, \- LM040\_02). 
On the other hand, the mass accretion rates for the AMC
cases are in the range 0.838 to 0.951.

For the LMC scheme, as the central hole becomes smaller, mass and angular
momentum accretion also become smaller with the standard grid. When
the central hole is large, a temporary accretion disk is formed and the
averaged mass accretion rate is close to the Hoyle \& Lyttleton
value. But for a small central hole an almost permanent accretion disk
is formed near the hole which blocks further accretion. However in the
fine grid case this decrease of the mass accretion is not found and the
existence of the accretion disk is temporary. On the other hand for the
AMC scheme, accretion disks are always transient and accretion rates
are similar.

This fact indicates that the conservation of angular momentum is very
important for the accretion problem. Since the results based on the
AMC scheme have less mesh dependency, they seem to be more
reliable. Thus only the results from the AMC scheme will be shown in
the rest of this section. 

\begin{figure}
\resizebox{0.49\hsize}{!}{\includegraphics{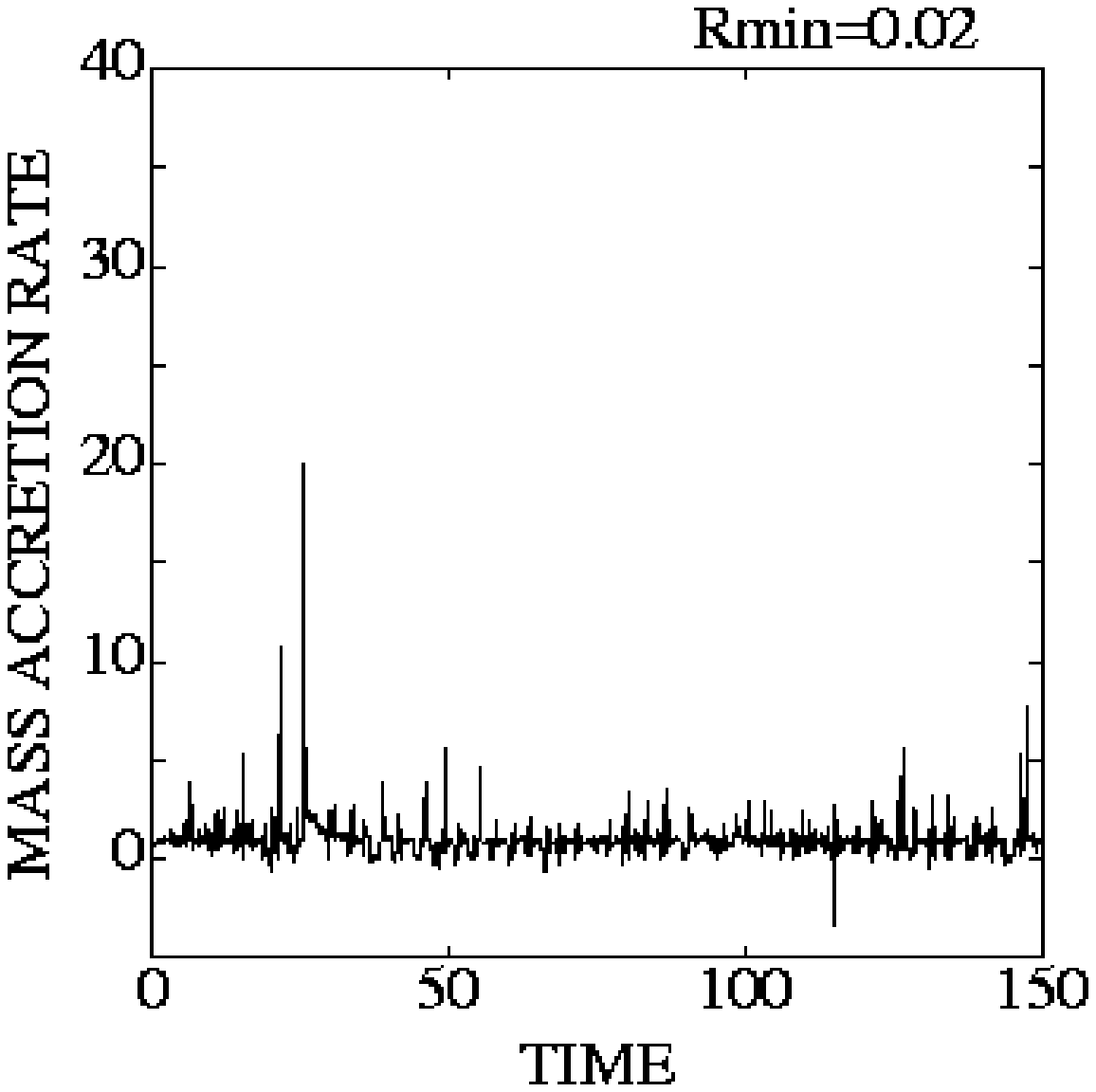}} \hfill
\resizebox{0.49\hsize}{!}{\includegraphics{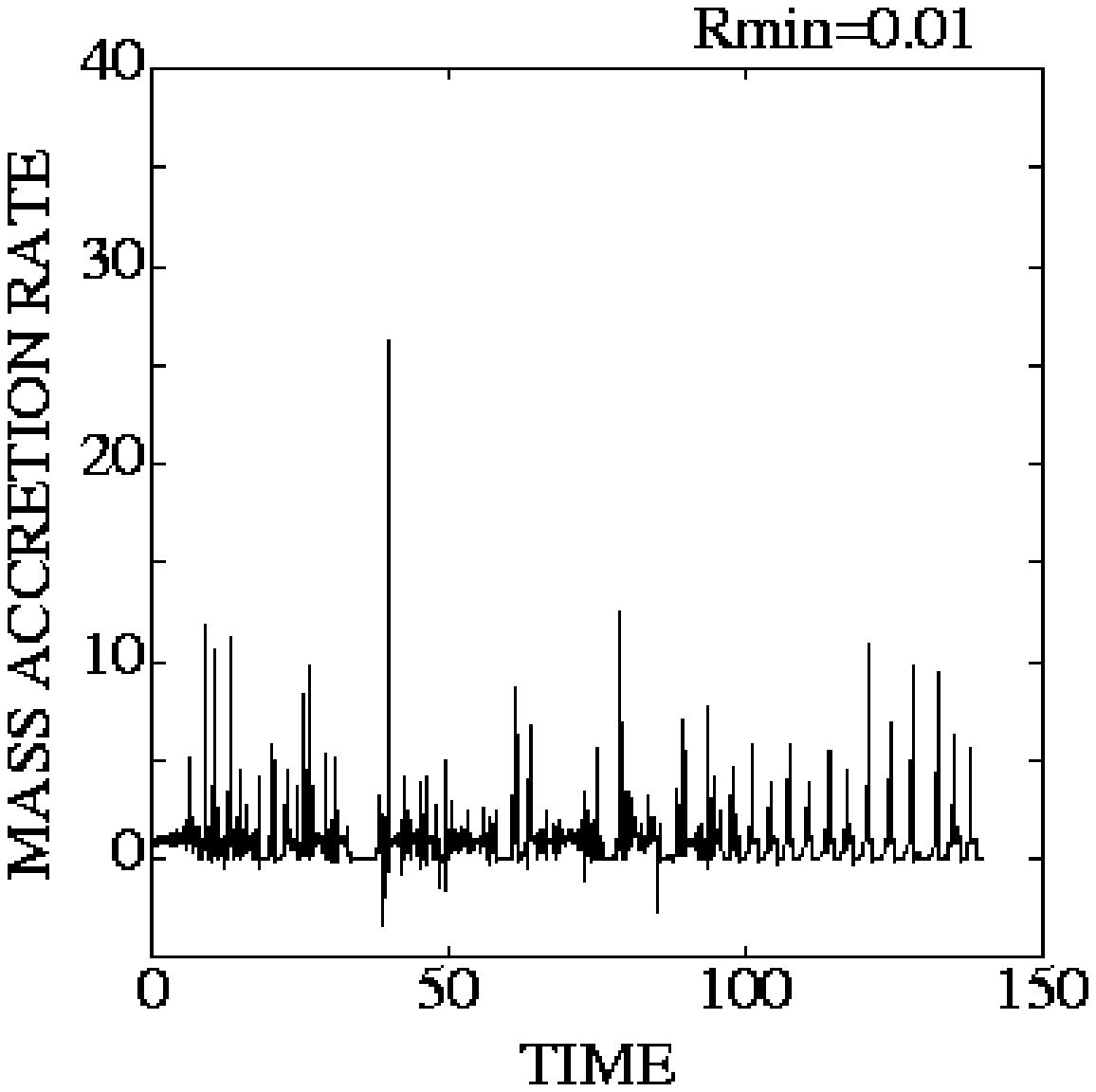}} \vspace{4mm}\\
\resizebox{0.49\hsize}{!}{\includegraphics{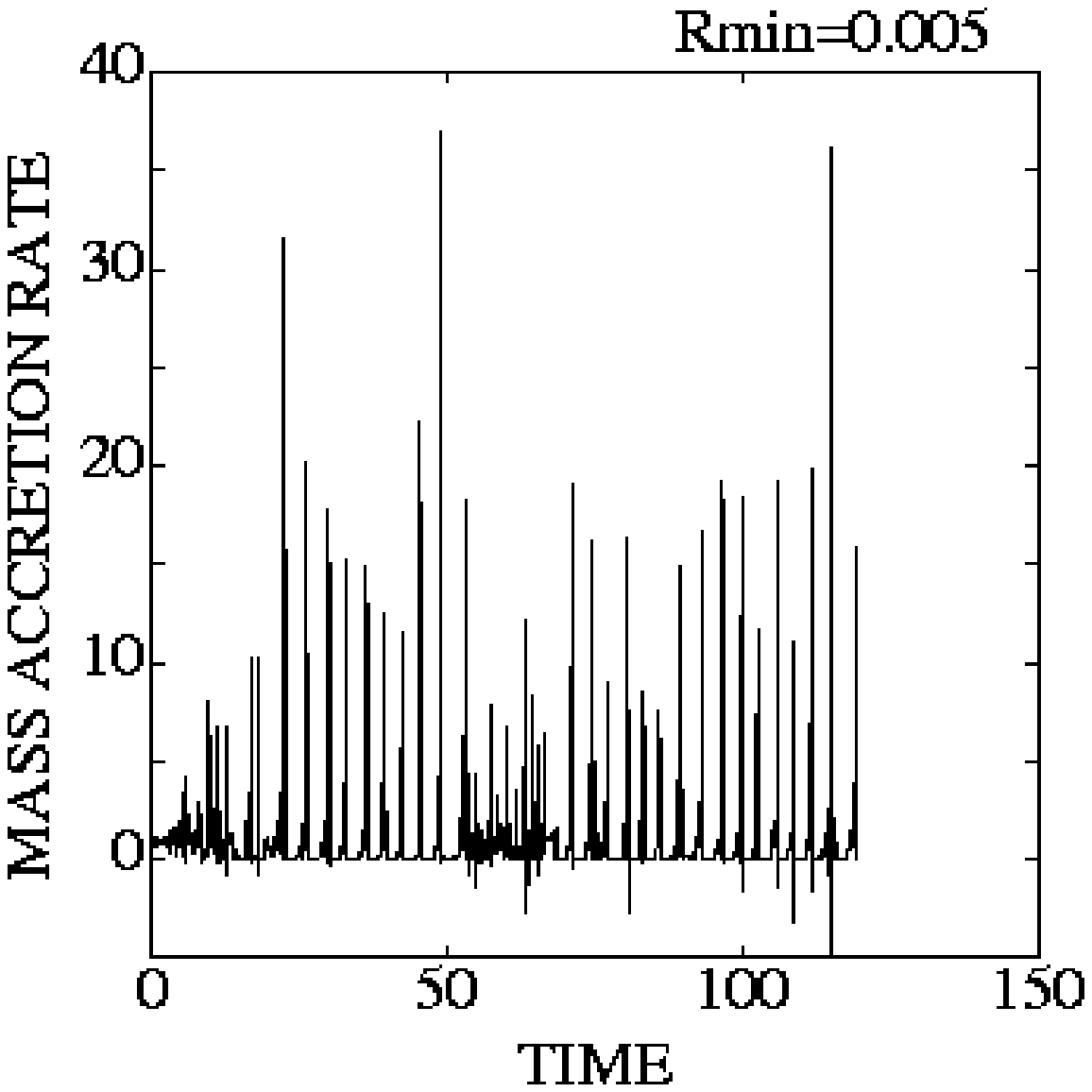}} \hfill
\resizebox{0.49\hsize}{!}{\includegraphics{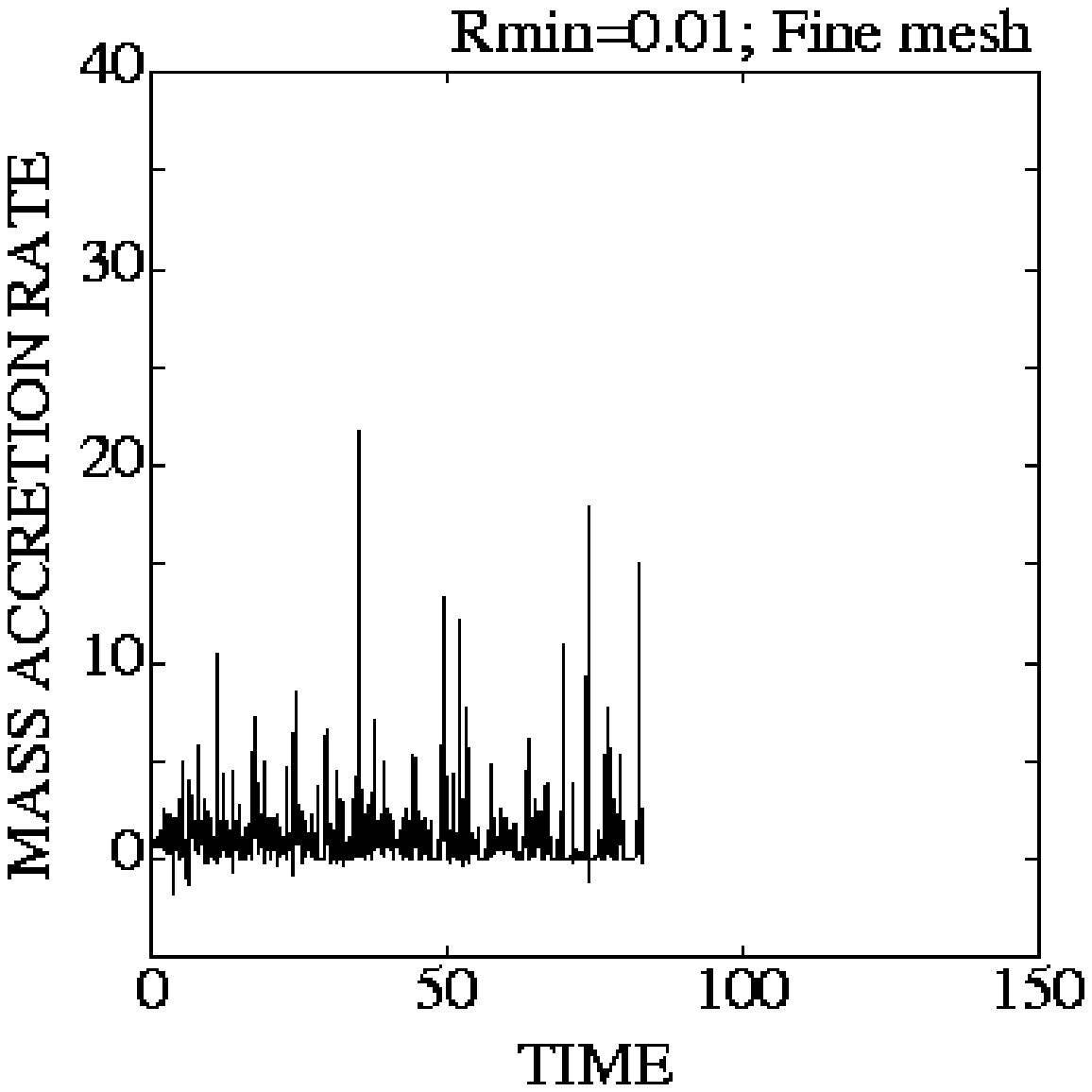}} \\
\caption{Time history of the mass accretion rate. $\cal{M}$=4.0; $R_{\rm
min}$=0.02, .01,
.005 and .01 fine grid, AMC scheme}
\label{am04002m}
\end{figure}
 
\begin{figure}
\resizebox{0.49\hsize}{!}{\includegraphics{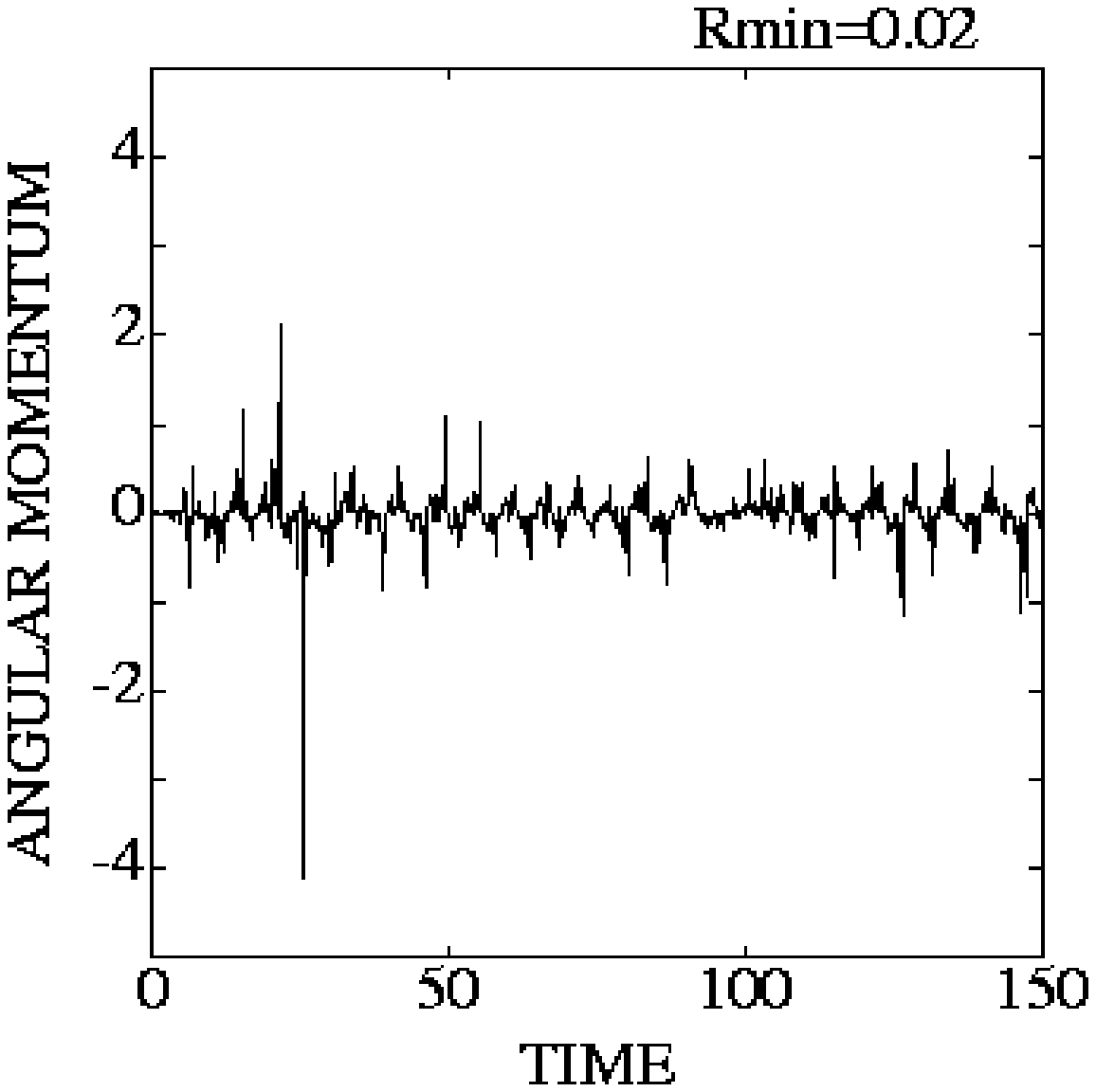}} \hfill
\resizebox{0.49\hsize}{!}{\includegraphics{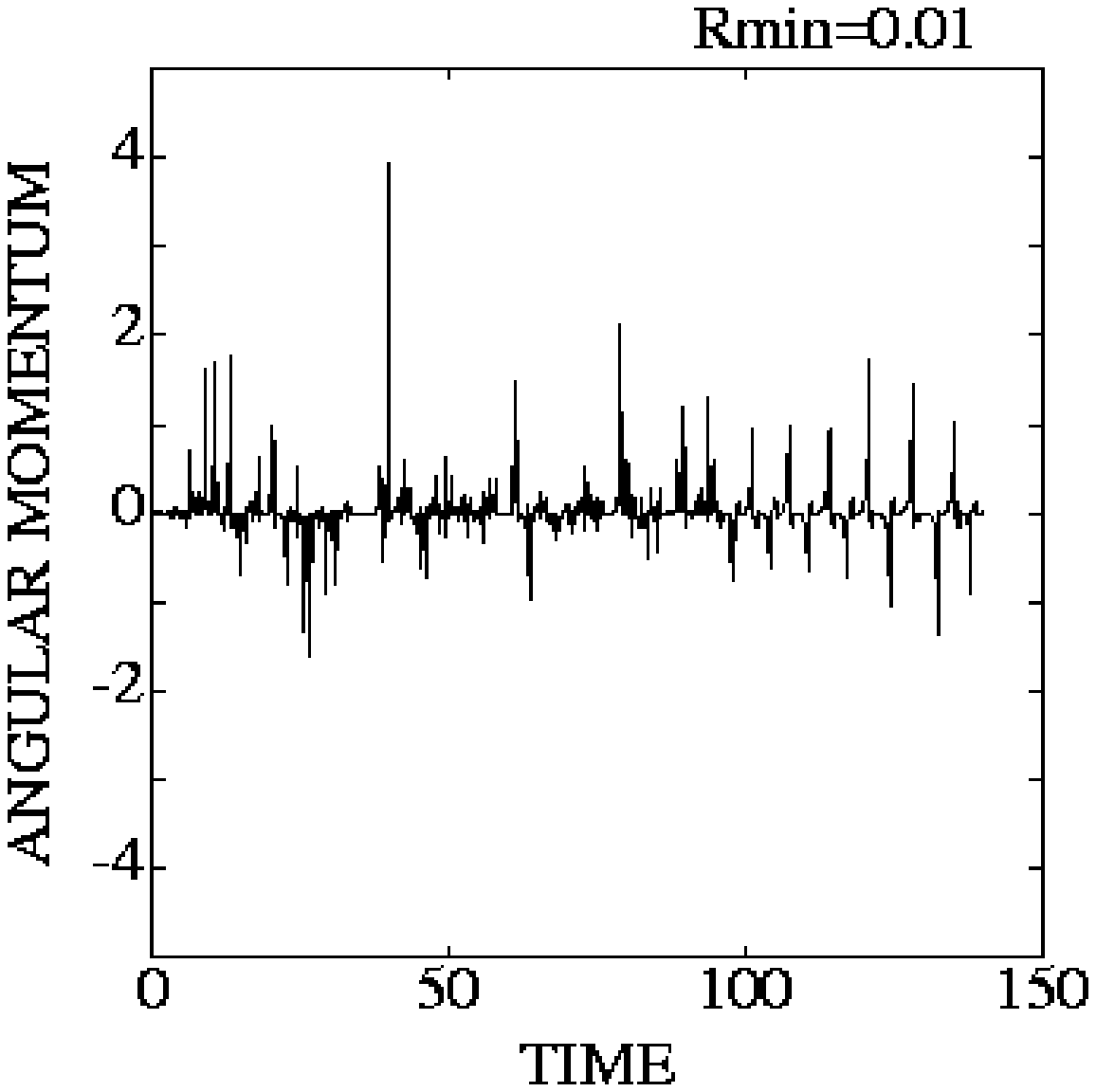}} \vspace{4mm} \\
\resizebox{0.49\hsize}{!}{\includegraphics{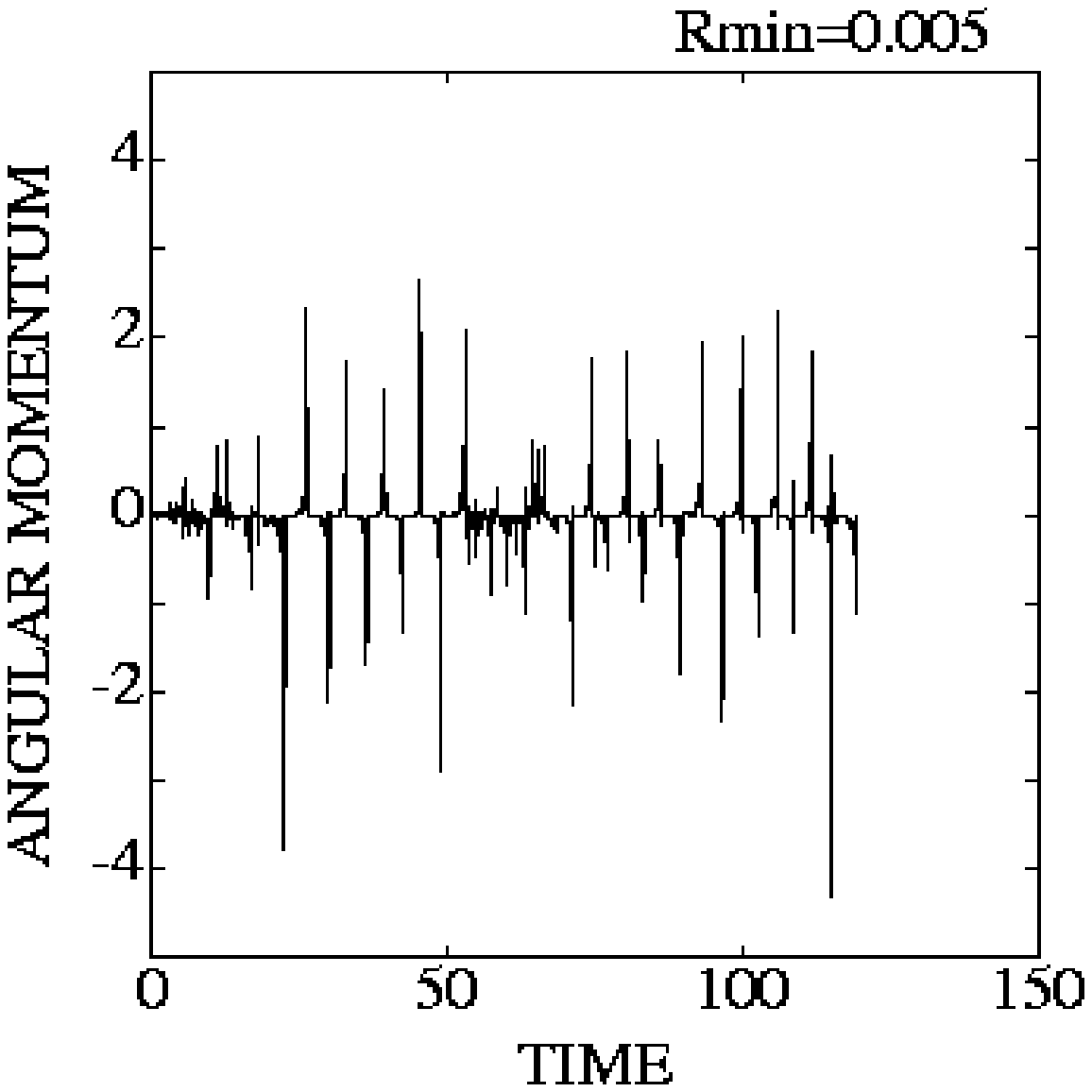}} \hfill
\resizebox{0.49\hsize}{!}{\includegraphics{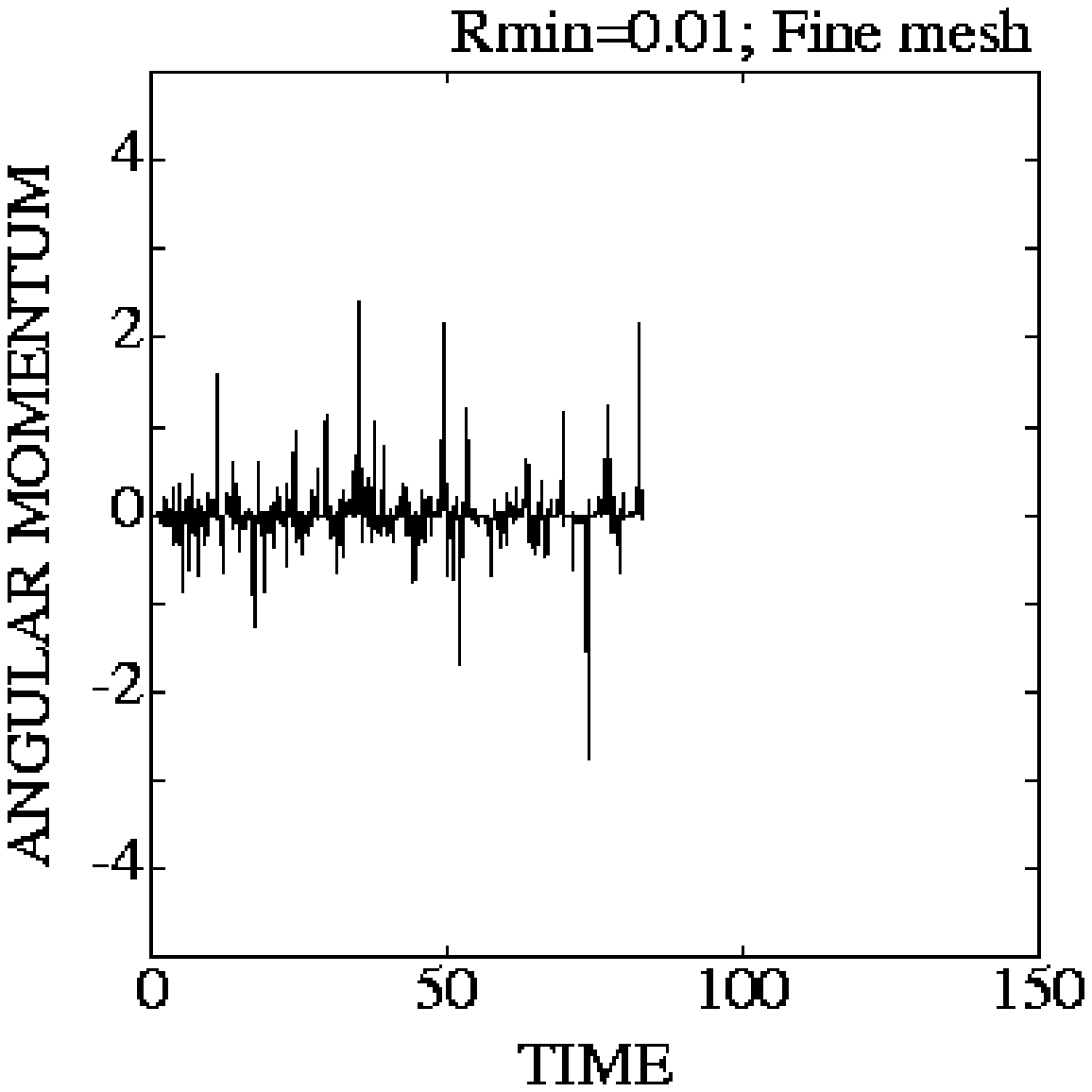}} \\
\caption{Time history of the angular momentum accretion rate.
$\cal{M}$=4.0; $R_{\rm min}$=0.02, .01,
.005 and .01 fine grid, AMC scheme}
\label{am04002a}
\end{figure}

\subsection{Occurrence of Oscillations and Formation of Accretion Disks}
The accretion column of the supersonic accretion flow is narrow and large
oscillations are found. In Figure~\ref{cont222s}, 
%,\ref{cont224s},\ref{cont226s},\ref{cont228s},\ref{cont230s},
the typical sequence of the swinging of the accretion column of case
AM040\-005 is shown. The oscillating accretion column swings over 180
degrees. This oscillation is similar to that shown in the computation
of Boffin and Anzer (\cite{bof}), but there is also an accretion disk
in our computation. The radius of the accretion disk is about 0.1 $R_{\rm a}$.

Enlarged views are shown in Fig.~\ref{cont221c}. The formation and
destruction of the accretion disk is clearly seen in these figures. An
anti-clockwise rotating accretion disk is formed at T=22. The
accreting matter falls from the upper-left direction through the
accretion column, thus this matter has anti-clockwise angular momentum
and it accelerates the disk. Then the column is \- pushed backward by
accreting matter from upstream. When the column moves behind the
object, the accreting matter has clockwise angular momentum
(T=22.2). This matter collides with the disk and destroys the disk
(T=22.4). Then a clockwise rotating disk is formed (T=22.6 and
following). As shown in Fig.~\ref{masbig}, mass and angular momentum
is accreted when the disk collapses. 

\begin{figure}
\resizebox{0.49\hsize}{!}{\includegraphics{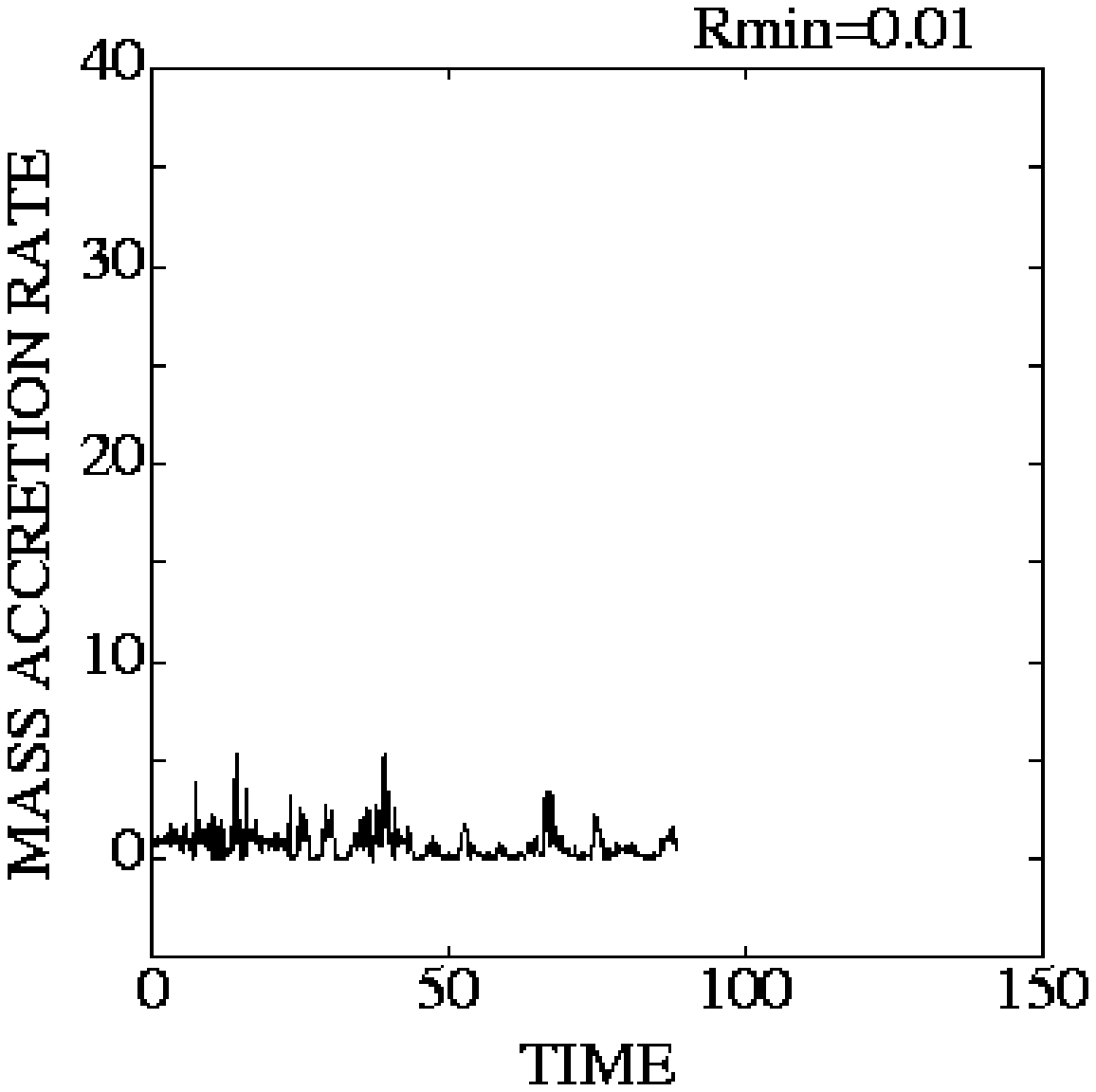}} \hfill
\resizebox{0.49\hsize}{!}{\includegraphics{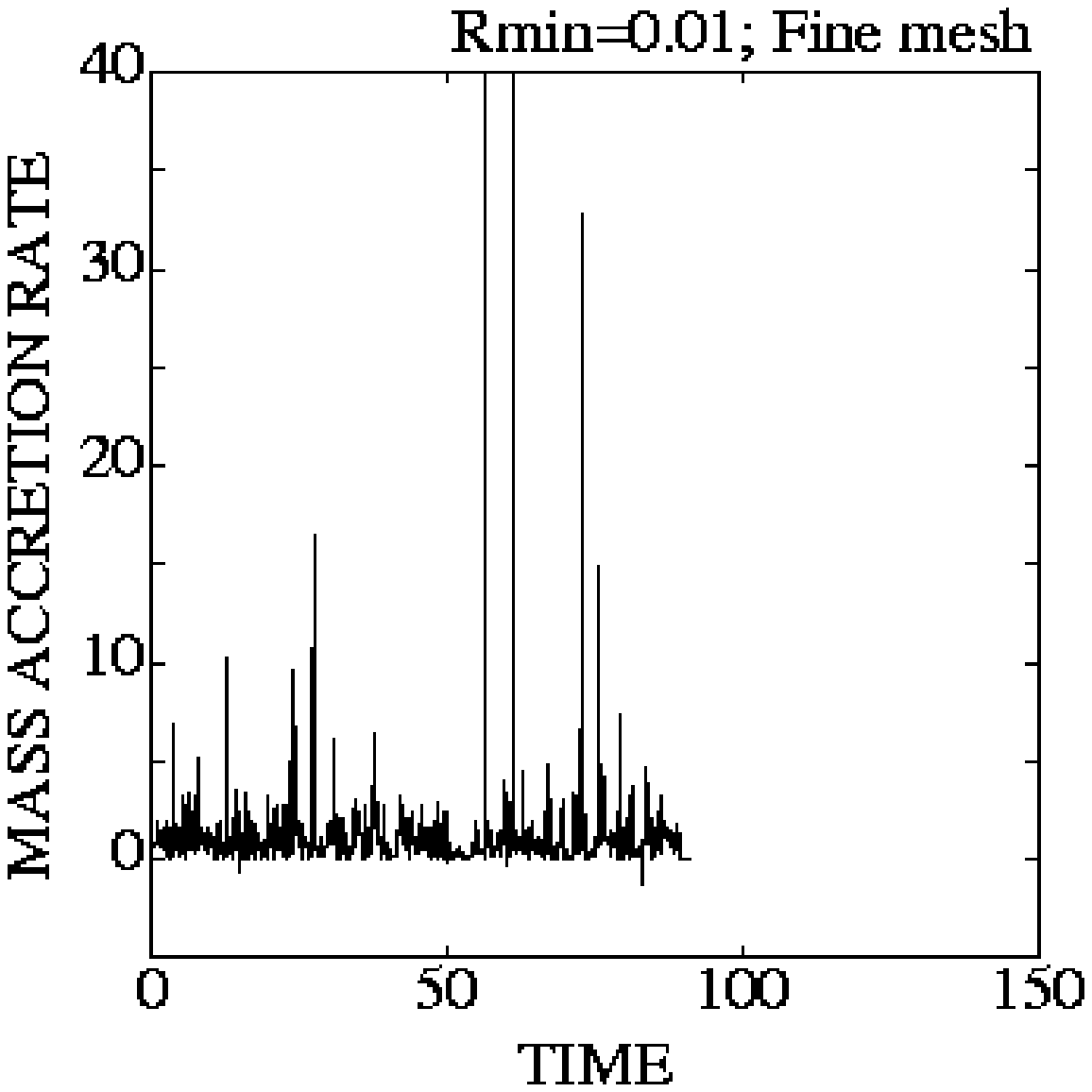}} \vspace{4mm} \\
\resizebox{0.49\hsize}{!}{\includegraphics{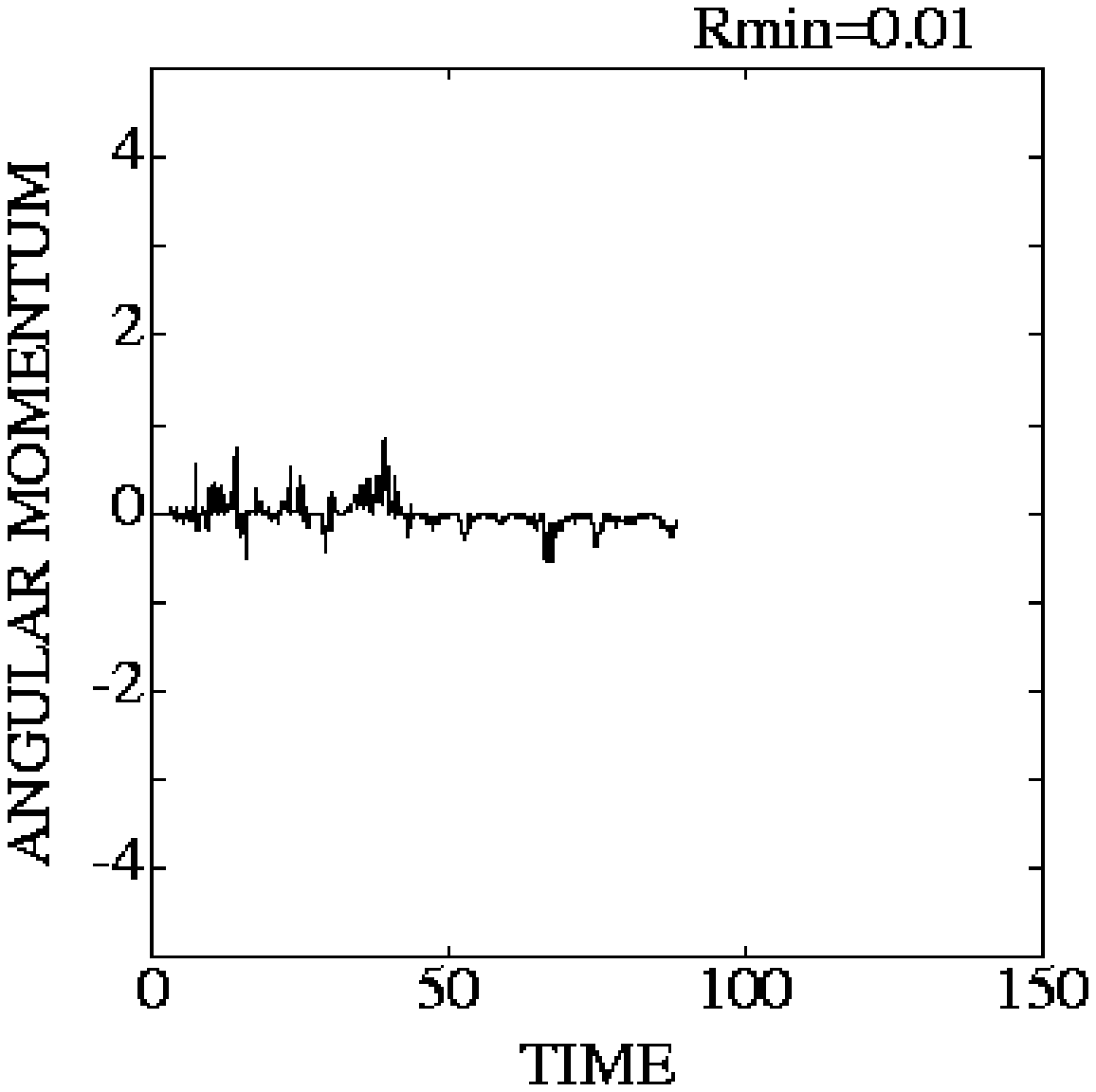}} \hfill
\resizebox{0.49\hsize}{!}{\includegraphics{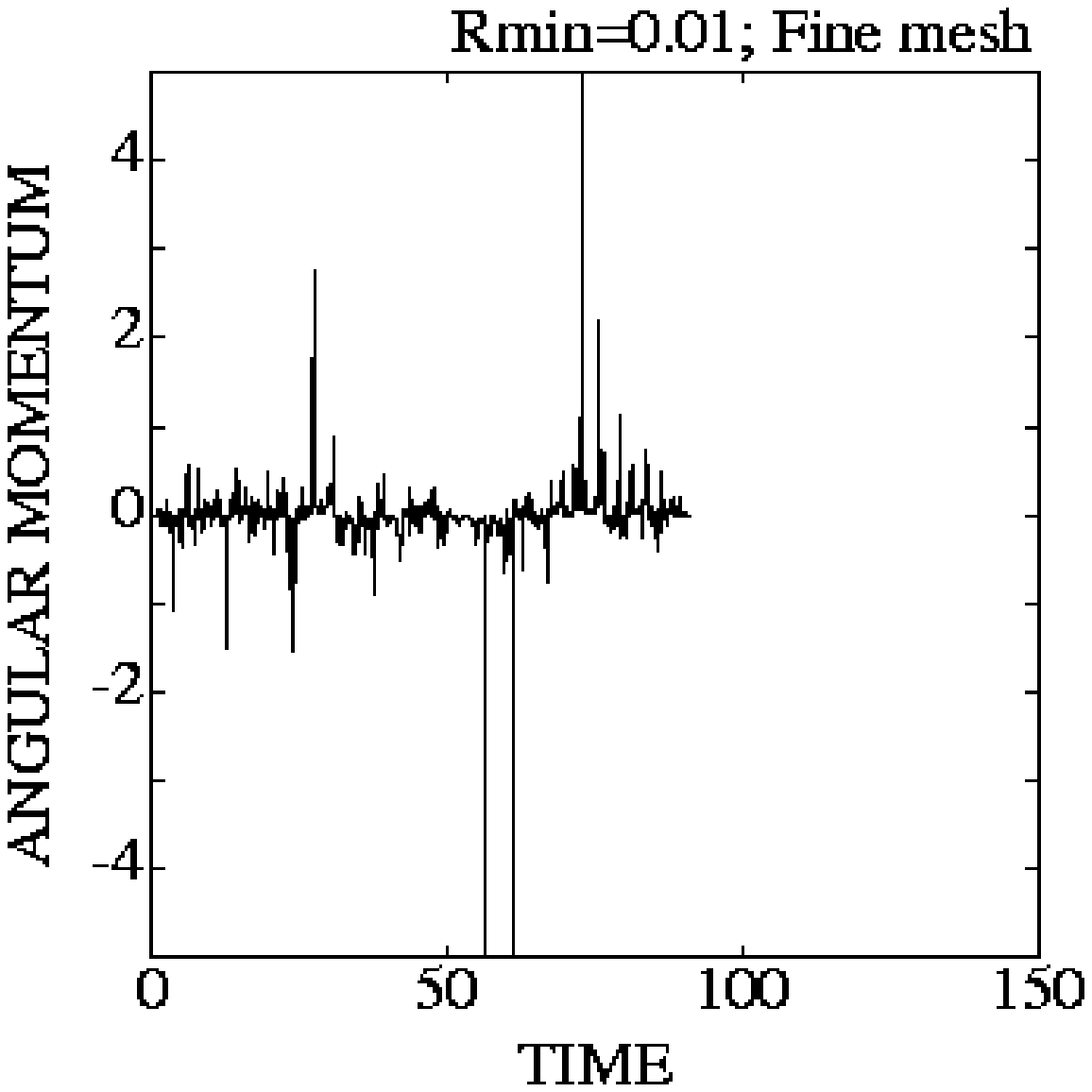}} \\
\caption{Time history of the mass accretion rate (upper frames) and 
angular momentum accretion rate (lower frames). $\cal{M}$=4.0 $R_{\rm
min}=0.01$.
The results from the standard grid (left) and from the fine grid (right)
are shown, LMC scheme}
\label{lm04001m}
 \end{figure}

The averaged mass accretion rate of this case is 0.855, but
the instantaneous rate is over 10 times larger than the Hoyle \& Lyttleton
estimate. The averaged momentum accretion is almost zero. However, a fair
amount of instantaneous  angular momentum accretion with altering direction
is also found. This can be understood as follows: the symmetric accretion
column cannot have angular momentum, but  the portions of an asymmetric
accretion column can have angular momentum. When the 
positive angular momentum portion falls downward, a positive accretion
disk is formed. This accretion disk loses its angular momentum and
accretes onto the object, when it collides with the following portion which
has negative angular momentum.

\subsection{History of Mass Accretion Rate and Angular Momentum Accretion}
The histories of mass accretion and angular momentum accretion of
representative cases from the AMC scheme are shown in
Fig.~\ref{am04002m} and \ref{am04002a}.  Those from the LMC scheme are
shown in Fig.~\ref{lm04001m}.

\subsubsection{Effects of $R_{\rm min}$ and Mesh size}
The histories of accretion rates for $\cal{M}$=4.0 from the AMC scheme with
different $R_{\rm min}$ and mesh size are shown in
Figs.~\ref{am04002m} and \ref{am04002a}.
As seen in those figures, the fluctuations of the accretion rate become
bigger with
smaller $R_{\rm min}$. 

\begin{figure}
\resizebox{\hsize}{!}{\includegraphics{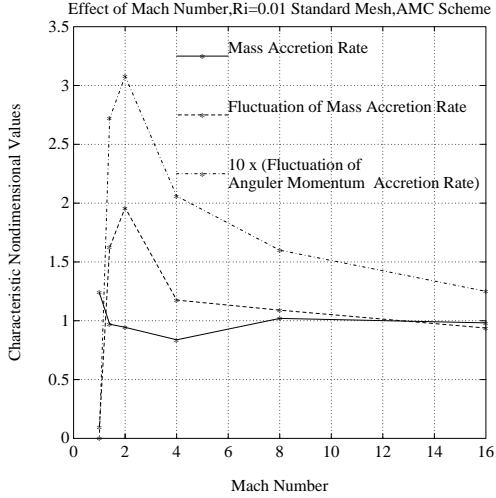}}
\caption{Mach number dependency of characteristic values for the AMC case with
 standard grid}
\label{rate}
\end{figure}

\begin{figure}
\resizebox{0.49\hsize}{!}{\includegraphics{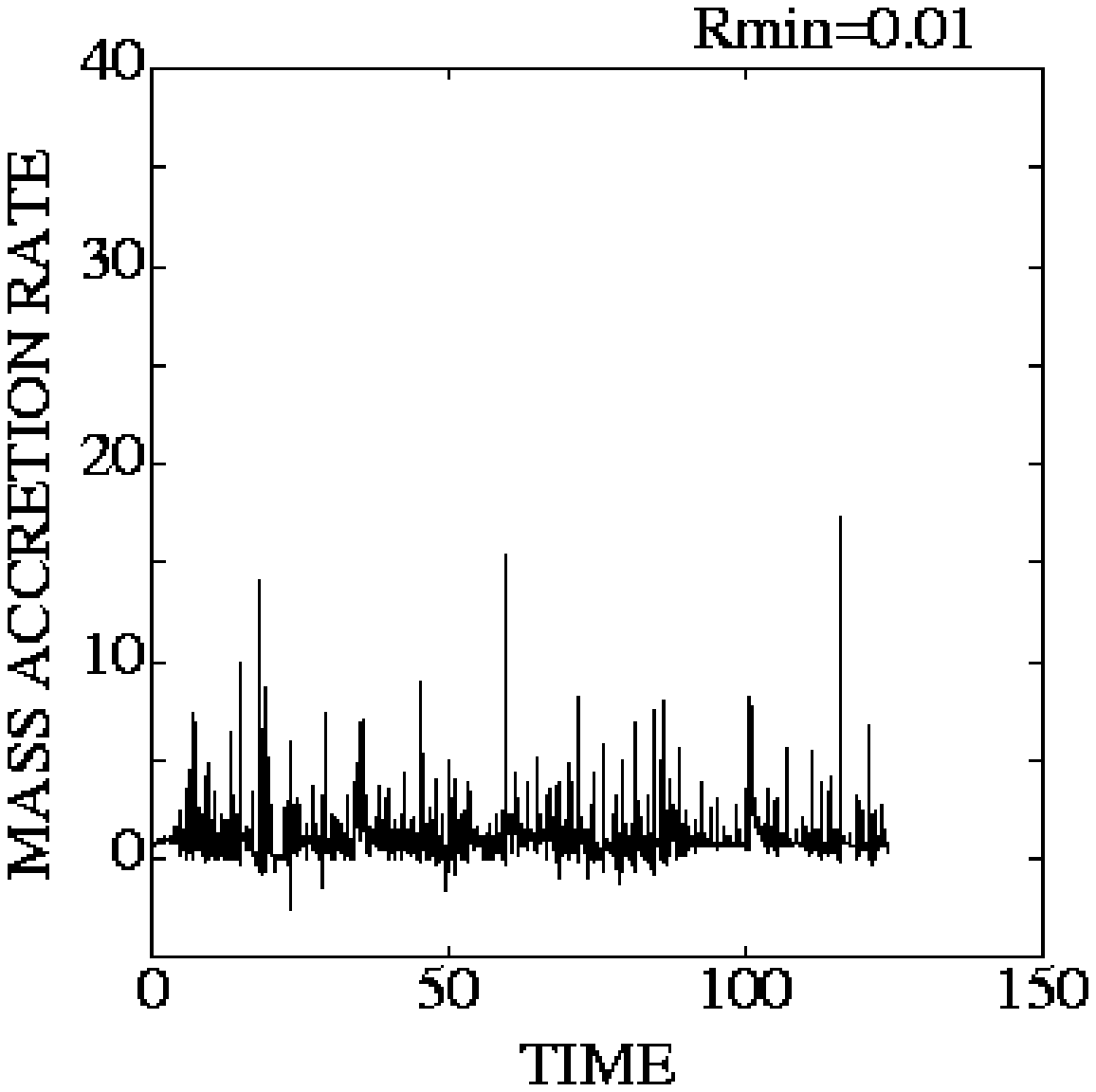}} \hfill
\resizebox{0.49\hsize}{!}{\includegraphics{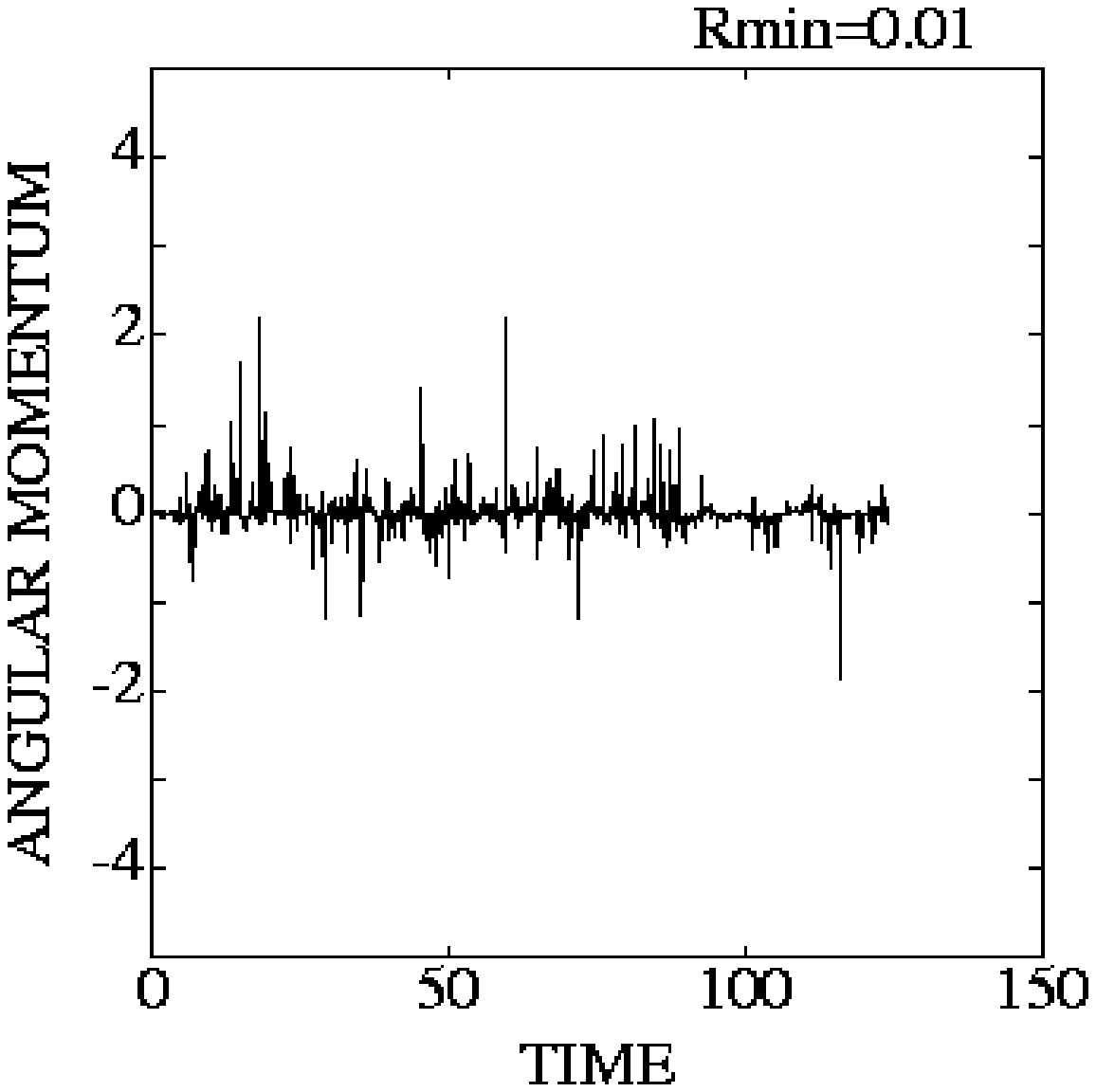}}
\caption{ Time history of the mass accretion rate (left) and
of the angular momentum accretion rate (right). $\cal{M}$=8.0 $R_{\rm
min=0.01}$}
\label{am08001m}
\end{figure}
 
We can see from table 2 and Fig. 4 that the mean values of the mass
accretion rate are quite similar for different values of $R_{min}$,
whereas the peak amplitudes of $\dot{M}$ increase with decreasing
$R_{min}$. This then implies that the accretion bursts are
correspondingly shorter for smaller values of $R_{min}$.
Although the results for the fine grid (Fig.~\ref{am04002m} \& \ref{am04002a})
have fluctuations of higher frequency and are more chaotic, they are
generally similar to those for the standard grid. The similarity
is also clearly seen in Table \ref{case}. All  values except 
$\dot{M}'$ are similar for different $R_{\rm min}$. The sequences of the
oscillations are also similar. The difference in $\dot{M}'$ can be explained
by the mechanism of the accretion disk destruction. The orbits of the
anti-clockwise rotating accretion disk in Fig.~\ref{cont221c} are
almost circular, but become elliptical when the accreting mater which
has clockwise angular momentum interacts with matter of the
anti-clockwise rotating disk (T=22.2 in Fig.~\ref{cont221c}). The
matter is taken away from the computational space when its elliptic
orbit touches the central hole. Thus, the bigger central hole absorbs the
matter over a longer period than the smaller hole. Therefore the
fluctuation of mass accretion becomes larger for a smaller hole.

On the other hand, the LMC scheme shows a different $R_{\rm min}$
dependence. For the standard grid, the mass accretion rate becomes
smaller as $R_{\rm min}$ decreases, as can be seen from the average
values of $\dot{M}$ which are given in table 2.. Because an almost
permanent accretion disk is formed it blocks further accretion. But
this permanent accretion disk is not found for the fine grid. In this case the
accretion rates show large fluctuations. The conservation of angular
momentum is only achieved to second order accuraey in the LMC
scheme. Thus, we believe that  the standard grid is not fine enough
for the LMC scheme to capture the accretion disk formation.

\subsubsection{Effects of the Mach Number} 
If the Mach number of the uniform flow is 1, the flow is perfectly steady, but
for larger Mach number it is always non-steady. Note that the case of
$\cal{M}$=1.4 is also non-steady and  shows small fluctuations from the
beginning. However, non-steadiness with large amplitudes occurs only after
40 time units.  The Mach number dependence of the averaged mass
accretion rate, the fluctuation of the mass accretion rate and the
fluctuation of the angular momentum accretion rate are summarized
in Fig.~\ref{rate}. These are the results from the AMC scheme with the
standard grid.  

As shown in Fig.~\ref{rate} the amplitude of the oscillations is largest when
the Mach number is 2, and the fluctuations become smaller as the Mach
number becomes greater than 4. This is also shown in the history of the
mass and angular momentum accretion of the Mach 8 case (case AM080\_01) in
Fig.~\ref{am08001m}.

\begin{figure*}
\resizebox{0.32\hsize}{!}{\includegraphics{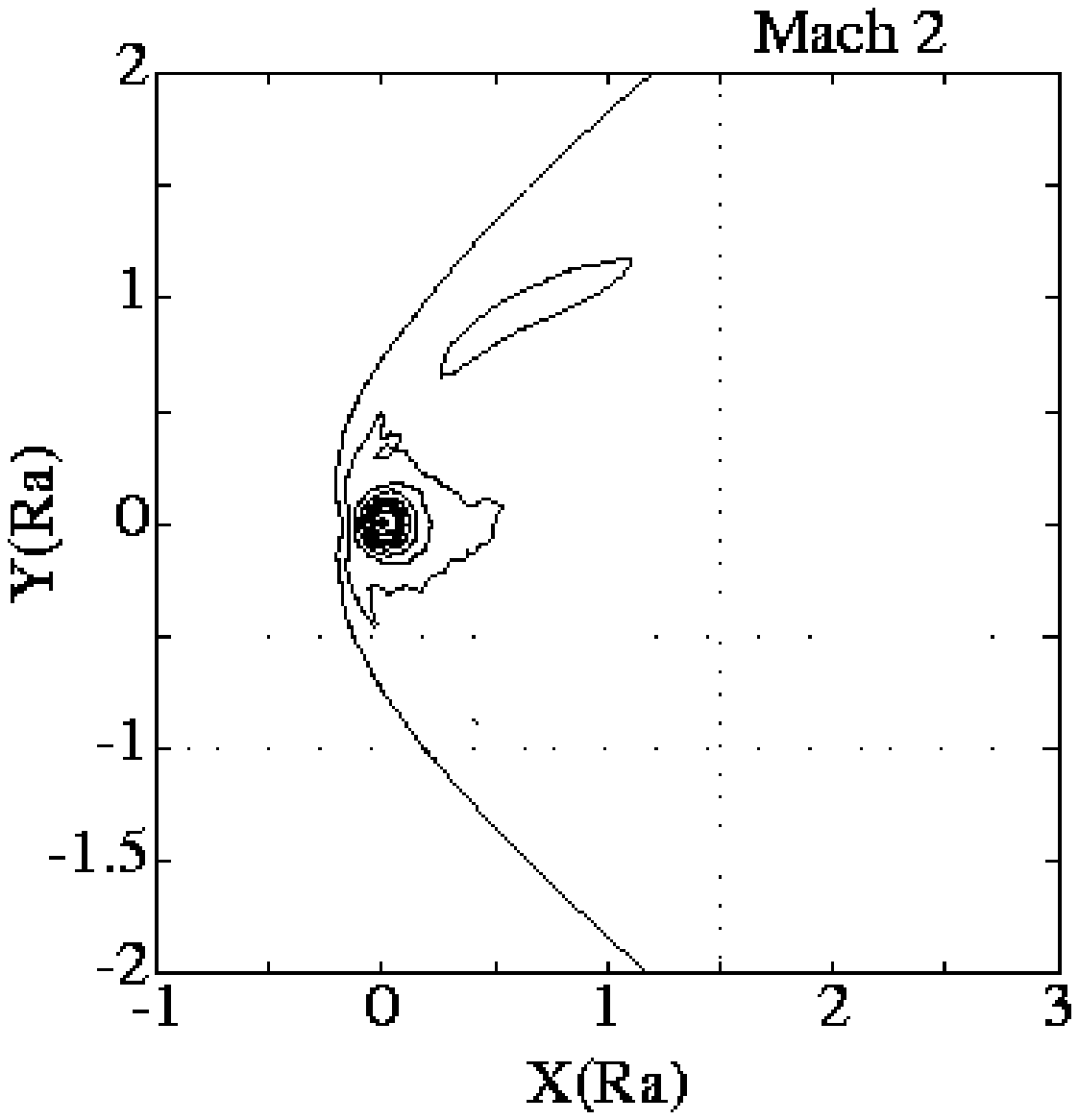}} \hfill
\resizebox{0.32\hsize}{!}{\includegraphics{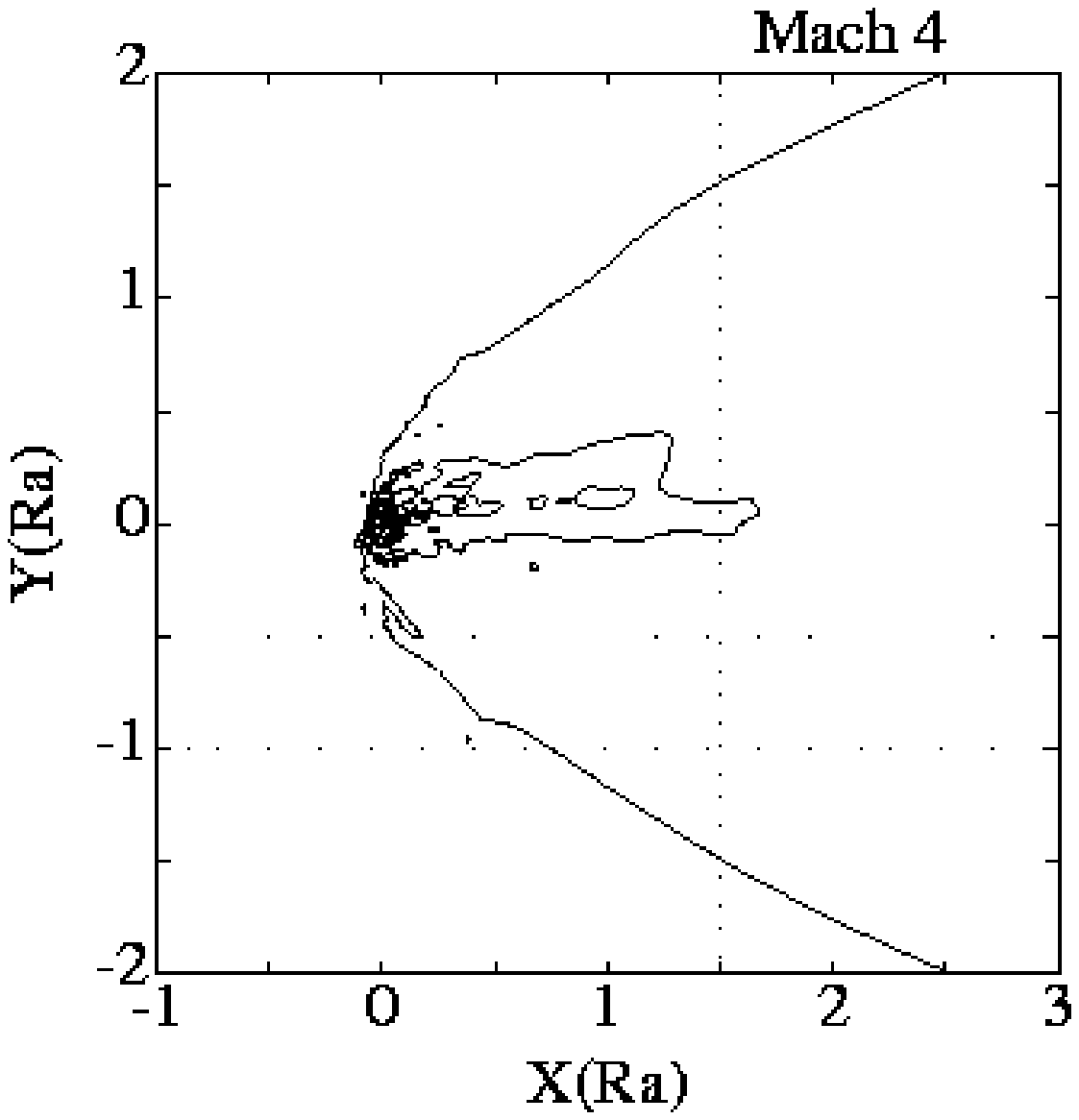}} \hfill
\resizebox{0.32\hsize}{!}{\includegraphics{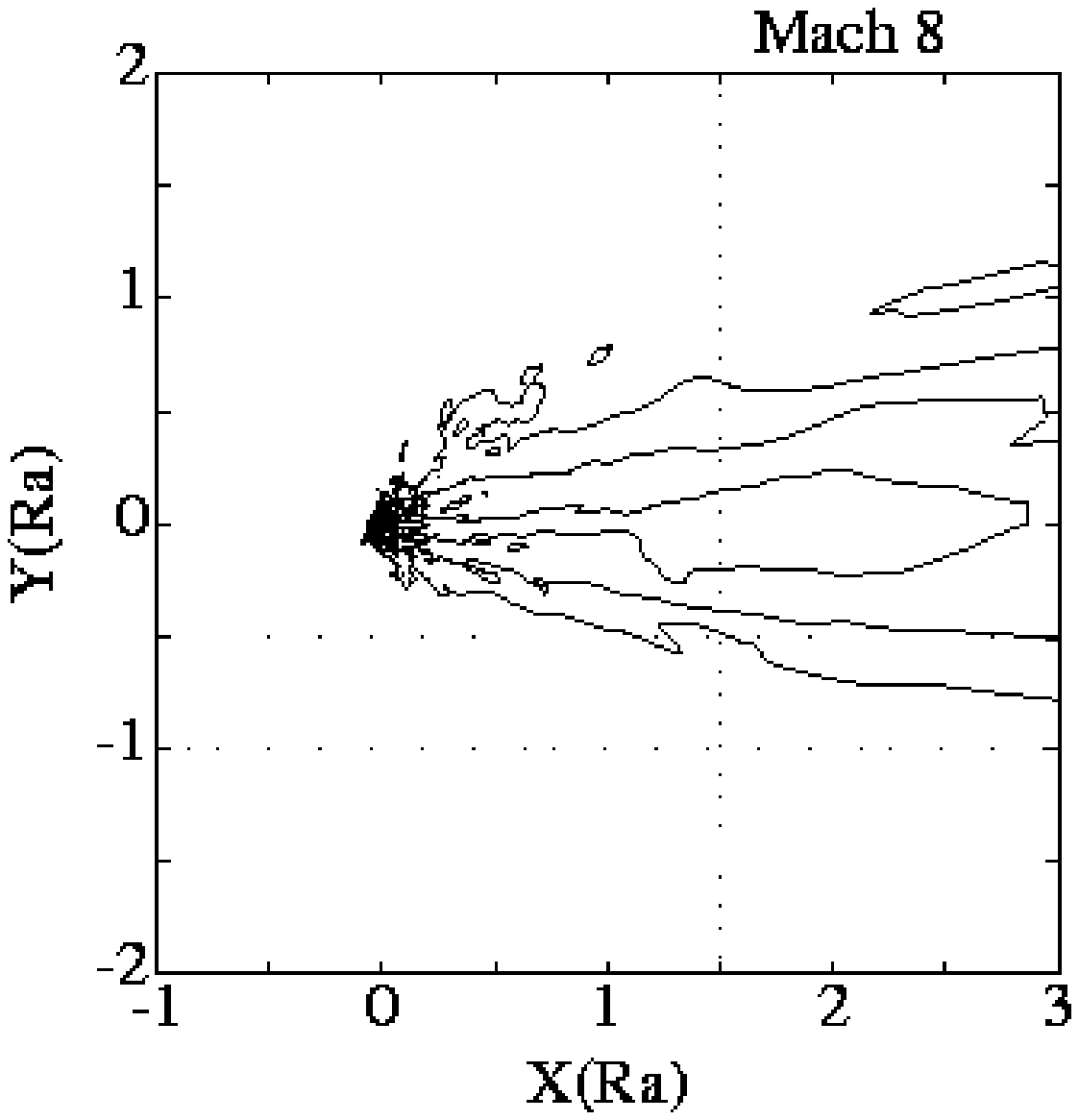}} 
\caption{Time averaged density contour. Rmin=0.01, $\cal{M}$=2,4 and 8. The
contours
are logarithmically spaced between 0 and 3, in steps of 0.3} 
\label{am02001av}
\end{figure*}

As the Mach number increases, the amplitudes of the accretion
column oscillation become smaller. This behavior is clearly shown
in the time averaged density contours of Fig.~\ref{am02001av}. In the time
averaged solution, the wiggles of the accretion column are smeared out
and the high density regions around the accreting object are seen. The
high density regions are concentrated in a narrow cone in the high
Mach number cases (see the ${\cal{M}}$=8 case in
Fig.~\ref{am02001av}). It is interesting that the transient accretion
disks are also seen as high density regions in the averaged solutions
and that averaged accretion columns look like bow shocks. 

\section{Summary and discussion}
Two dimensional planar wind accretion flows of a supersonic isothermal gas
were  analyzed numerically. The computational efficiency was improved
as compared to earlier calculations by using an accurate local time
stepping and a new type of upwind scheme.

Comparing our new results with earlier calculations one realises that
2D flows with $\gamma$ close to unity have so far not been studied
very systematically. Matsuda et al. (1992) presented one case with
$\gamma = 1.05$ and one with $\gamma = 1.005$. Their resolution near
the central object was much coarser than our resolution. They had also
obtained flip-flop configurations, but the instability took much
longer to develop. Their amplitudes were smaller and the widths of the
accretion columns considerably larger. Boffin and Anzer (1994) also
calculated one model with $\gamma = 1.1$ which has a narrow
oscillating accretion column, but again a low amplitude. 3D flows with
$\gamma = 1.01$ have been systematically studied by Ruffert (
1996). In these flows the accretion regions are always much wider and
do not show the systematic oscillations of entire accretion region,
although the accretion itself is non-steady. There differences between
3D and 2D flows are quite general and independent of the value of $\gamma$.

We also have developed a numerical scheme which exactly conserves
angular momentum (AMC scheme) and compared its performance to the
earlier linear momentum conserving (LMC) scheme. We found that the
results obtained with our LMC scheme depend very strongly on the
resolution of the grid used, whereas the AMC scheme gives almost no
difference for the two types of resolution considered here. Therefore
we feel that the AMC calculations are much more reliable and all the
results presented in the previous section are based on this scheme. The
fact that our AMC and LMC results differ by large amounts suggests to
us that many of the earlier calculations which basically conserve
linear momentum should be taken with caution. This will be
particularly important for the temporal behavior of the accretion of
angular momentum $\dot{J}$ (t).

Our calculations of supersonic flows show that in all cases the
accretion of both mass and angular momentum is very
erratic. It occurs in very short intervals and at extremely high
rates. This behavior can be explained by the formation of Keplerian disks
near the inner boundary. If the specific angular momentum of the
infalling material is larger than that of the Kepler orbit at the
innermost radius then this material cannot accrete.

If such high angular momentum material is flowing in long enough, a
disk will form which blocks further accretion very
efficiently. Material with very low angular momentum or with opposite
rotation falling in during a subsequent phase interacts with the disk
and can destroy it. This will lead to a burst of the accreted mass and
angular momentum. After such a burst the process can be repeated and a
new disk will form. Our calculations indicate that reversals of the
disk rotation are quite common.

For the modeling of X-ray binaries fed by wind accretion the
fluctuations of $\dot{J}$ are of major importance. They can be \- brought
into relation with the observed spin-up and spin-down of these X-ray
pulsars; see Anzer \& B\"orner (\cite{anz}). In their investigation they
showed that the random fluctuations calculated by Ruffert (\cite{Rufa}) for
3d models were by a factor $\approx10$ too low in order to explain the
pulse period variations observed in the source Vela X-1. 
However our new calculations give variations of $\dot{J}$ which are
substantially larger than those found by Ruffert. We have obtained
typical values of RMS $(\dot{J})$ of the order of 0.2 in our dimensionless
units (see Table
\ref{case}). This result can also be formulated as:
%\begin{equation}
\begin{eqnarray*}
RMS (\dot{J}) & = & 0.2~\rho _\infty ~{v^2_\infty} ~ R_{\rm a}^2  \\ 
%RMS (\dot{J}) 
& = & 0.1~ \dot{M}_{HL}~{v_\infty}~ R_{\rm a}  \\
%RMS (\dot{J}) = 
& \approx & 0.1~{v_\infty}~R_{\rm a}
%\end{equation}
\end{eqnarray*}
since $\dot{M}$ is typically of the order unity. On the other hand
Ruffert (\cite{Rufd}) gives RMS(j)=0.01${v_\infty} R_{\rm a}$ for
$\gamma=1.01$ and RMS(j) = 0.03${v_\infty}R_{\rm a}$ for $\gamma = 5/3$. Taking
into account that RMS($\dot {J}$)=RMS ($\overline{\dot{M}}$j) and
$\overline{\dot{M}}$ $\approx1$ we have RMS ($\dot{J}$) =
(0.01-0.03)${v_\infty}R_{\rm a}$. Therefore our values for the
fluctuations area factor 3 to 10 larger than those of Ruffert's 3D
calculations. Therefore on the basis of our calculations one might
conclude that the observed period fluctuations could in principle, be
caused by random fluctuations of $\dot{J}$. But the amplitudes are
only marginally large enough and any slight reduction of the efficency
would rule out this interpretation. There is in particular the aspect
that our calculations are two--dimensional whereas the real flows are
three--dimensional and the difference in amplitudes between 2D and 3D
flows could be sufficiently large to make the described interpretation
invalid. To really answer this question requires full 3D computations,
taking angular momentum conservation into account.

\begin{acknowledgements}
HB acknowledges the support of PPARC grant GR/K 94157 as well as a
Royal Society travel grant. TM thanks Cardiff University of Wales for
its hospitality. We also thank the referee, Max Ruffert for his fast
response and his helpful suggestions.
\end{acknowledgements}


\begin{thebibliography}{99}
\bibitem[1995] {anz} Anzer, U., B\"orner, G., 1995, A\&A, 299, 62
\bibitem[1997] {ben} Benensohn, J.S., Lamb, D.Q., Taam, R.E., 1997, ApJ,
478, 123
\bibitem[1979] {bis} Bisnovatyi-Kogan, G.S., Kazhdan, Ya.M., Klypin, A.A.,
Lustkii, A.E., Shakura, N.I., 1979, SvA, 23, 201
\bibitem[1994] {bof} Boffin, H.M.J., Anzer, U., 1994, A\&A, 284, 1026
\bibitem[1952] {bon52} Bondi, H., 1952, MNRAS, 112, 195
\bibitem[1944] {bon44} Bondi, H., Hoyle, F., 1944, MNRAS, 104, 273
\bibitem[1982] {Osher} Chakravarthy, S.R., Osher, S., 1982, AIAA  Paper
82-0975
\bibitem[1988] {fry} Fryxell, B.A., Taam, R.E., 1988, ApJ, 335, 862
\bibitem[1989] {Haenel} H\"anel, D., Schwane, R.,  AIAA Paper 89-0274, 1989
\bibitem[1989] {ho} Ho, C., Taam, R.E., Fryxell, B.A., Matsuda, T., Koide,
H., Shima, E., 1989, MNRAS, 238, 1447
\bibitem[1939] {Hoy} Hoyle, F., Lyttleton, R.A., 1939, {\it
Proc.Cam.Phil.Soc.}, 35,405
\bibitem[1971] {Hun71} Hunt, R., 1971, MNRAS, 154, 141
\bibitem[1979] {Hun79} Hunt, R., 1979, MNRAS, 188, 83
\bibitem[1993] {Ish} Ishii, T., Matsuda, T., Shima, E., Livio, M., Anzer, U.,
B\"orner, G., 1993, ApJ, 404, 706
\bibitem[1993] {sfs} Jounouchi, T., Kitagawa, I., Sakasita, S., Yasuhara,
M., 1993, {\it
Proceedings of 7th CFD Symposium,} in Japanese
\bibitem[1991] {Koi} Koide, H., Matsuda, T., Shima, E., 1991, MNRAS, 252, 473
\bibitem[1993] {Liou} Liou, M.S., Steffen, C.J., 1993, J. Comp. Phys., 107, 23
\bibitem[1991] {Liv} Livio, M., Soker, N., Matsuda, T., Anzer, U., 1991,
MNRAS, 253, 633 
\bibitem[1987] {Mat87} Matsuda, T., Inoue, M., Sawada, K., 1987, MNRAS, 226, 785
\bibitem[1991] {Mat91} Matsuda, T., Sekino, N., Sawada, K., Shima, E.,
Livio, M.,
Anzer, U., B\"oner, G., 1991, A\&A, 248, 301
\bibitem[1992] {Mat92} Matsuda, T., Ishii, T., Sekino, N., Sawada, K.,
Shima, E.,
Livio, M., Anzer, U., 1992, MNRAS 255, 183.
\bibitem[1981] {Roe} Roe, P.L, 1981, J.Comp.Phys., 43, 357
\bibitem[1994a] {Rufa} Ruffert, M., 1994a, ApJ, 427, 342
\bibitem[1994b] {Rufb} Ruffert, M., 1994b, A\&AS, 106, 505
\bibitem[1995] {Rufc} Ruffert, M., 1995, A\&AS, 113, 133
\bibitem[1996] {Rufd} Ruffert, M., 1996, A\&A, 311, 817
\bibitem[1994] {Rufe} Ruffert, M., Arnett, D., 1994, ApJ, 427, 351
\bibitem[1989] {Saw} Sawada, K., Matsuda, T., Anzer U., B\"orner, G., Livio, M.,
1989, A\&A, 231, 263
\bibitem[1997] {SIMA} Shima, E. , Jounouchi, T., 1997, {\it
NAL-SP34,Proceedings of 14th NAL
symposium on Aircraft Computational Aerodynamics,} p.7
\bibitem[1985] {Shi} Shima, E., Matsuda, T., Takeda, H., Sawada, K., 1985,
MNRAS 217, 367
\bibitem[1981] {SW} Steger, J.L., Warming, R.F., 1981, J.Comp.Phys., 40, 263
\bibitem[1988] {Taa} Taam, R.E., Fryxell, B.A., 1988, ApJ, 327, L73
\bibitem[1982] {VL} van Leer, B., 1982, {\it Lecture Note in Physics}, 170, 507
\bibitem[1994] {Wada} Wada, Y. , Liou, M.S., 1994, AIAA Paper 94-0083
\end{thebibliography}
\end{document}